\documentclass{aa}  

\usepackage{txfonts}

\usepackage{graphicx,longtable,lscape,stfloats}
\usepackage{natbib}
\bibpunct{(}{)}{;}{a}{}{,} 
\newcommand{\rsun}{R$_{\odot}$}

\newcommand{\ltsima} {$\; \buildrel < \over \sim \;$}  
\newcommand{\gtsima} {$\; \buildrel > \over \sim \;$}  
\newcommand{\lta} {\lower.5ex\hbox{\ltsima}}  
\newcommand{\gta} {\lower.5ex\hbox{\gtsima}}

\usepackage{color}
\usepackage[normalem]{ulem}
\usepackage{amstext}

\begin{document}

\title{Contribution of polar plumes to fast solar wind}
\subtitle{} \titlerunning{Polar plumes and fast wind}
\authorrunning{Zangrilli and Giordano}

\author{L.~Zangrilli \and S.~M.~Giordano}

\institute {Istituto Nazionale di Astrofisica (INAF) - Osservatorio Astrofisico di Torino, via Osservatorio 20, 10025 Pino Torinese, Italy}

\offprints{luca.zangrilli@inaf.it}

 
\abstract
    {Several physical properties of solar polar plumes have been identified by different published studies, however such studies are rare and sometimes in disagreement.}
    {The purpose of the present work is to analyze a set of SOHO/UVCS data dedicated to the observation of plumes and to obtain a picture of the physical properties of plumes in the intermediate solar corona through a self-consistent analysis.}
    {We applied the Doppler Dimming technique to data acquired by SOHO/UVCS in April 1996, which was during the very early phases of the mission. From this we derived outflow speeds and electron densities. We used SOHO/LASCO images as context data in order to better identify plume and interplume regions in the UVCS field of view.}
    {The results we obtain demonstrate that in three cases out of four plumes expand with outflow speeds comparable to those of interplumes, and in a single case with lower speeds. We estimate that the contribution of plumes to the wind coming from the solar poles is about 20\%, and that different plumes provide a different contribution, possibly according to different stages of their evolution.}
    {We conclude that plumes are not static structures, and that they contribute significantly to the wind coming from the solar poles.}

\maketitle

\section{Introduction}
\label{sec:intro}

Polar plumes are bright structures observed in the polar coronal holes of the Sun, which extend from the solar surface far into the corona, and with a geometry similar to open magnetic field lines (see, e.g., \citealt{DeForest_2001b}). Several studies published in the literature have established that plumes are relatively slowly expanding structures, characterized by higher densities and lower temperatures than the surrounding interplume plasma. The present status of knowledge about plumes has been reviewed by \cite{Wilhelm2011a} and \cite{Poletto2015a}; a comprehensive review of the physics of coronal holes and of their sub--structures is given by \cite{Cranmer2009a}.

The role of plumes and interplumes as contributors to the fast solar wind has been greatly debated, both structures being recognized in the past as possible fast wind sources \citep{Gabriel2003a, Gabriel2005a}. Some authors assign a marginal role to plumes, giving more importance to the interplume plasma as the origin of the fast wind. The analysis of observations from the SOHO/UVCS (SOlar and Heliospheric Observatory/UltraViolet Coronagraphic Spectrometer) and SOHO/SUMER (Solar UV Measurements of Emitted Radiation) spectrometers \citep{Kohl1997a, Cranmer1999a,wilhelm_k_1995a}, and the application of the Doppler Dimming technique (see, e.g., \citealt{Noci1987a}), allowed us to estimate the outflow speed in plumes and interplumes in the intermediate corona. From an analysis of UVCS observations at 1.70~\rsun\, of heliocentric distance, \cite{Giordano2000a}  concluded that plume data are well reproduced by a static model, while interplume plasma is dynamically expanding. Moreover, in \cite{Teriaca2003a}, the analysis of a combination of SUMER and UVCS data led to the determination of  plume dynamics from 1.05 up to 2.0~\rsun, and to the deduction that plumes are essentially static structures. From stereoscopic reconstructions based on STEREO/SECCHI (Solar TErrestrial RElations Observatory/Sun Earth Connection Coronal and Heliospheric Investigation) observations, \cite{Feng_2009a} investigated the 3D geometry of plumes and, combining it with SUMER data, from a Doppler analysis deduced a maximum outflow speed for the O~{\sc{vi}} ions on the order of 10~km/s, at distances not higher then 200~Mm from the solar limb, corresponding to about 1.3~\rsun.

However, different authors came to different conclusions. For example, from Doppler shift measurements taken by Hinode/EIS (Extreme-ultraviolet Imaging Spectrometer), \cite{Fu2014a} deduced a steady acceleration in plumes at heights lower than 1.05~\rsun, and concluded that their contribution to the fast wind must be not negligible. A scenario of dynamically expanding plasma in plumes is also implied by \cite{Raouafi2007a}, who found that the shape of the observed line profiles and the line intensities are better reproduced in plumes by a low-speed regime at low altitudes, increasing with height, and reaching interplume values above 3-4~\rsun. For these reasons we could say that, up to now, a definite consensus on fast wind sources seems not to have been reached.

As we have seen, estimates of the outflow speed of plumes in the intermediate corona deduced by many authors (see, e.g., \citealt{Wilhelm2011a}, \citealt{Fu2014a}) are rare and sometimes in disagreement. Hence it is of interest to obtain additional measurements in this interval of heliocentric distances. A good opportunity is offered by the analysis of UVCS spectroscopic data, acquired during a campaign of observations of polar plumes in the north polar coronal hole carried out in 1996, during the minimum of solar activity (see Sect.~\ref{sec:data} for a detailed description). Although partial results from the data set we are considering in the present study have been already published in \cite{Giordano2000a}, they only analyzed data at 1.7~\rsun. Hence the full range of altitudes still has to be explored, and our purpose is to extend the Doppler Dimming analysis over a larger range of heliocentric distances. Our approach is to find models for plumes and interplumes, capable of reproducing the H~{\sc{i}}-Lyman $\alpha$ and O~{\sc{vi}} line intensities from UVCS observations.

The plume lifetime covers a time interval from hours to several days, and it has been pointed out that during their life plumes can disappear and then show up again at the same position \citep{Lamy_P_1997a, DeForest2001a}. Moreover, plumes show variability with time on timescales that depend on the observational spatial scales \citep{DeForest1997a}. In particular, plumes appear stationary for at least one day when observed at scales larger than $10^{\prime\prime}$ in the low corona.

In the present study, we assume that plumes are homogeneous. Over the past years, evidence has grown suggesting that plumes might not be homogeneous. This has been demonstrated, for example, by  \cite{Woo_r_2006a}, who showed how the structures observable in coronal holes have a filamentary aspect on spatial scales down to $\sim 31$ km.

We will consider possible effects of plume variability with time and inhomogeneity in space, when discussing the results from our analysis of the data in the present paper. Our observations have been acquired during an overall time interval of a couple of days, and the observing time at a single altitude in the corona is on the order of a few hours (see Sect. \ref{sec:data}), leading to the problem of plume identification when comparing results obtained at different heliocentric distances.

The paper is organized as follows: in Sect.~\ref{sec:data} we describe the data, in Sect.~\ref{sec:analysis} we summarize the procedure by which we inferred the physical parameters of plumes and interplumes. Section~\ref{sec:results} describes the outcomes from the application of the Doppler Dimming analysis, which are then discussed in Sect.~\ref{sec:discussion}, where a summary of the main results is given and conclusions are drawn.

\section{Data}
\label{sec:data}

\subsection{SoHO/UVCS data and data analysis}
\label{sec:uvcs_data}

We analyzed SoHO/UVCS (\citealt{Kohl1995a}) coronal spectra acquired above the north polar coronal hole over the time interval 6-9 April 1996, taken in the UVCS spectral channels of the O~{\sc{vi}} doublet lines at 1031.9 and 1037.6~\AA,\, and the H~{\sc{i}}~Ly$\alpha$ line at 1215.7~\AA. The spectral and spatial information were binned over two-by-two detector pixels, corresponding to a spectral resolution of $0.2$~\AA\ and $0.28$~\AA\ per bin, for the O~{\sc{vi}} and H~{\sc{i}}~Ly$\alpha$ channels, respectively, and to a spatial resolution of $14^{\prime\prime}$ per bin in both channels. In the plane of the sky, $14^{\prime\prime}$ correspond to about 0.014~\rsun.

The coronal instantaneous field of view imaged onto the entrance slits of the O~{\sc{vi}} and H~{\sc{i}}~Ly$\alpha$ channels covered different heliocentric distances, from 1.38 to 2.53~\rsun, along the radial through the north pole. We refer to the altitude in the corona of the instantaneous field of view as the slit position or altitude. Slit widths of $49~{\rm \mu m}$ were used in both channels to have a good spectral resolution and photon flux, corresponding to an instantaneous field of view $14^{\prime\prime}$ wide, normal to the slit direction.

The exposure time of each acquired frame was 600~s, for the overall 72~hrs of observations. Because of the faint emission from coronal holes, we limited our analysis to slit positions below 1.9~\rsun, and increased the signal to noise ratio by grouping the observations of three nearby altitudes, and by summing the photon counts of the corresponding spectral and spatial bins, so averaging data over $42^{\prime\prime}$ in the direction perpendicular to the slit. A further spatial rebin along the slit resulted in a spatial resolution of $28^{\prime\prime}$ per bin in that direction. The total exposure times, the mean observation altitudes, and the dates of observation of the resulting spectra, are summarized in Table~\ref{tab:01}.

\begin{table}
  \caption{Parameters of the analyzed spectral data. These were derived from UVCS observations of the plume study campaign in the period 6-9 April 1996.}
\centering
\begin{tabular}{l l l c c}
\hline\hline
ID & Exposures  & $r_{mirror}$ & Total exp. time & Date \\
~  &     ~      & $(R_\odot)$ & (s)             & ~    \\
\hline
\hline
0  &   0-19     & 1.375    & 12000              & 6 Apr.\\
1  &  20-44     & 1.442    & 15000              & 6 Apr. \\
2  &  45-74     & 1.524    & 18000              & 6 Apr. \\
3  &  75-109    & 1.620    & 21000              & 6 Apr. \\
4  & 110-149    & 1.730    & 24000              & 7 Apr. \\
5  & 150-194    & 1.855    & 27000              & 7 Apr. \\
\hline
\end{tabular}
\label{tab:01}
\end{table}

We studied the spectral lines of the O~{\sc{vi}} doublet and the redundant H~{\sc{i}}~Ly$\alpha$, which are recorded by the O~{\sc{vi}} detector and cover the same instantaneous field of view in the corona. Data were analyzed using the UVCS/Data Analysis Software, which takes care of wavelength and radiometric calibration. The UVCS radiometric calibration and its evolution in time is discussed by \cite{Gardner2002a}. The stray-light contribution has been subtracted according to the results obtained by \cite{tesi_phd_silvio}.

\subsection{Detector bias response correction}
The identification of plumes and interplumes in UVCS data could be affected by instrumental biases introduced by the detector. We constructed a flat field describing the uneven response of the detector by averaging a large number of coronal spectra at different latitudes and distances, based on the hypothesis that individual coronal structures are smoothed out by the sum of many different and unrelated contributions. Suitable observations are those of the synoptic program. During our observations no synoptic run was taken, hence we had to rely on the synoptic data taken during the three days preceding our observations. The synoptic program consists of observations routinely covering the intermediate corona at different position angles and heliocentric distances. The resulting intensity profiles along the slit for the two UVCS spectroscopic channels have been corrected for an intensity decrease of the corona in the instantaneous field of view, where different heliocentric distances are sampled. The correction for the coronal intensity decrease was done by normalizing the profiles by themselves after applying a spatial smoothing.

\begin{figure}
  \centering
  \includegraphics[width=4.5cm]{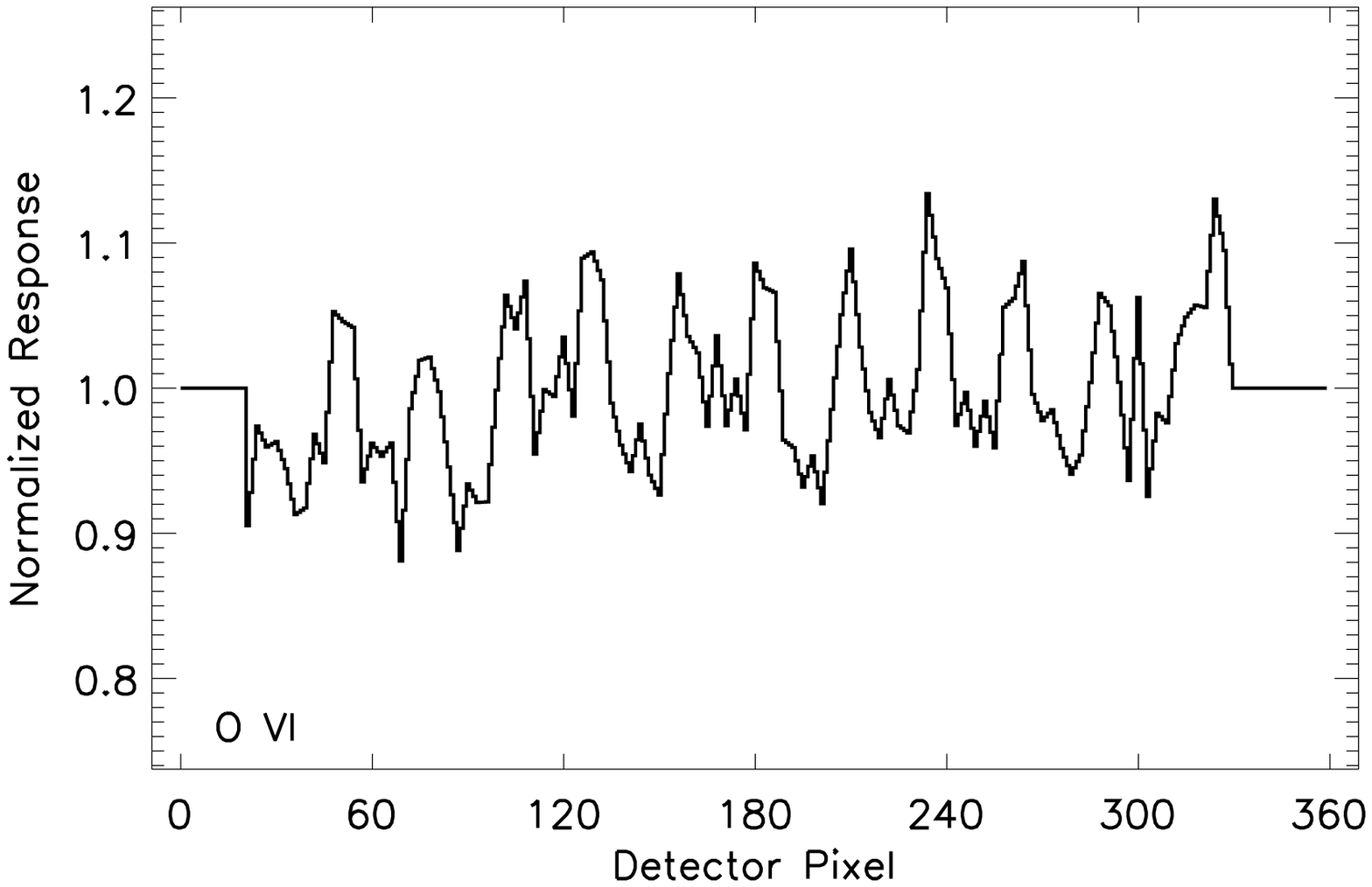}\includegraphics[width=4.5cm]{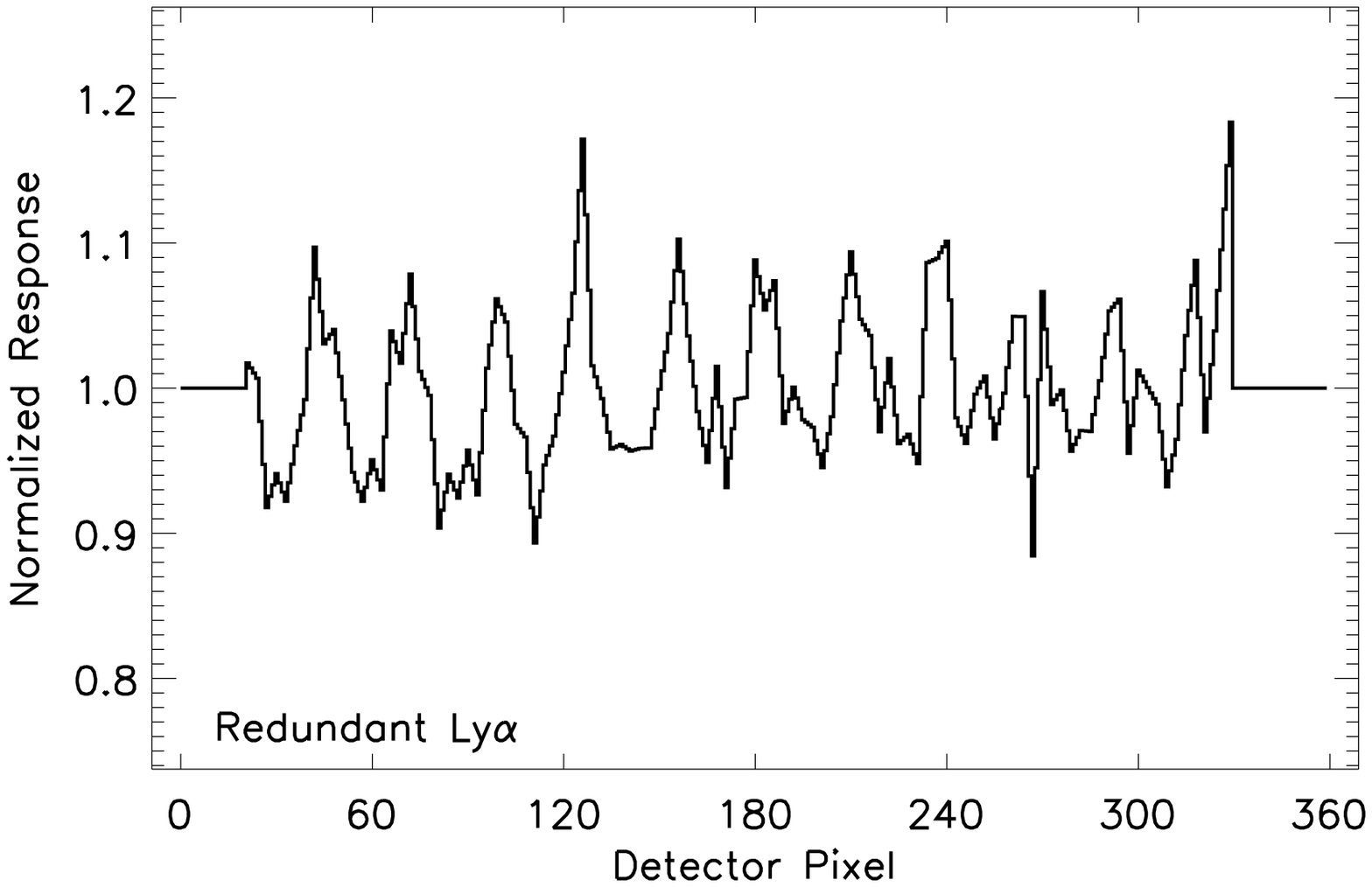}\\
  \includegraphics[width=4.5cm]{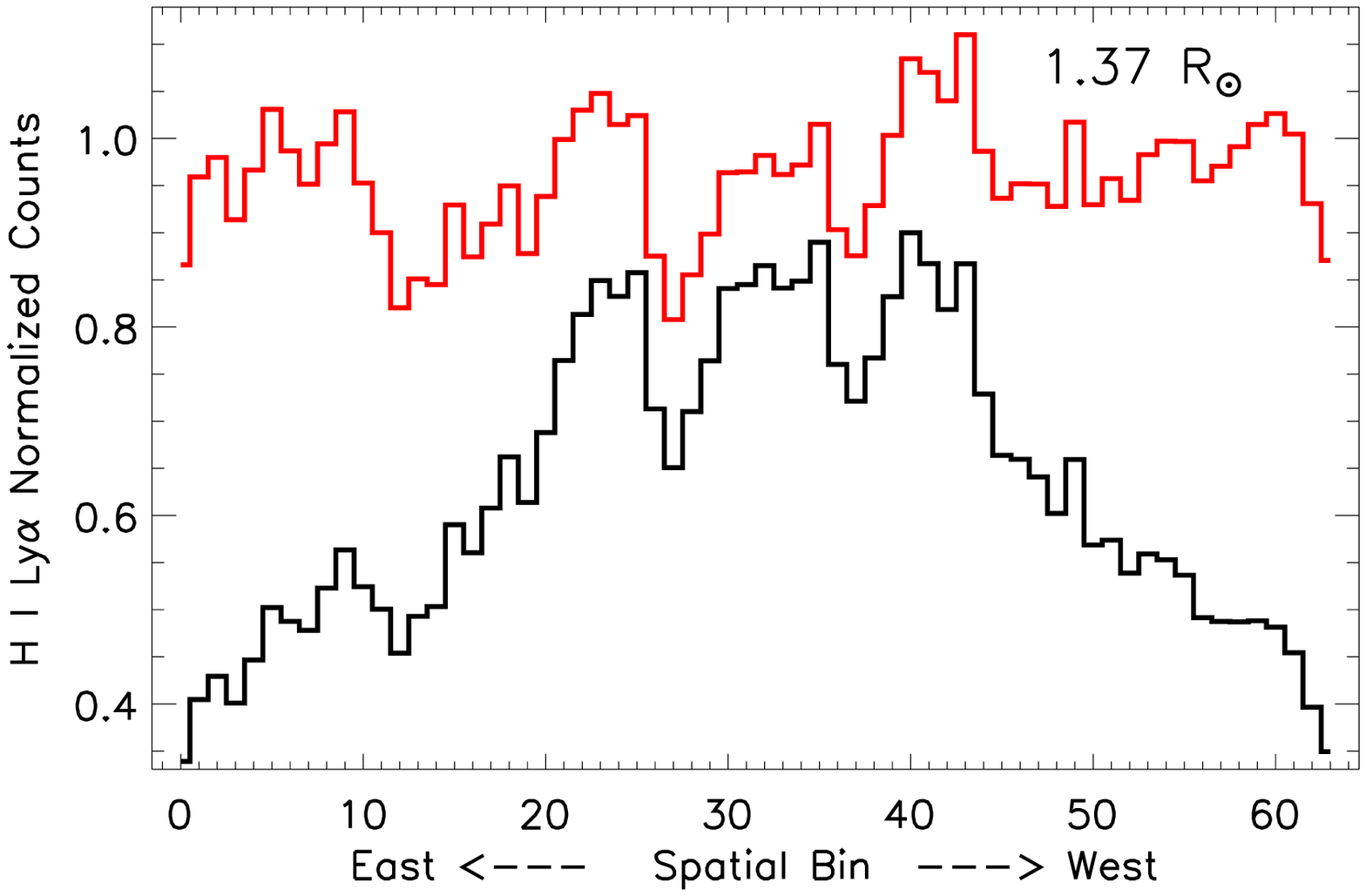}

  \caption{Flat field response of the O~{\sc{vi}} channel along the UVCS slit in the O~{\sc{vi}} doublet (top-left panel) and redundant H~{\sc{i}}~Ly$\alpha$ (top-right panel) spectral intervals. An example of bias-corrected data is shown in the bottom panel, where the original data are represented by a black curve and the corrected data are shown with a red curve. The data corrected for the detector flat field response have also been normalized to the value measured at the pole, only in this figure for graphical reasons, in order to account for the coronal intensity decrease, when moving from the slit center to parts of the corona at higher distances (see text for correction description).}
  \label{fig:uvcs_flat}
\end{figure}

The analysis described above revealed a periodic spatial variation in the response of the detector of the O~{\sc{vi}} channel, with an amplitude of about 10\% and a period of eight to ten pixels, as shown in Fig,~\ref{fig:uvcs_flat}, in the two parts where the primary and redundant spectra are sampled.

\begin{figure*}
  \centering
  \includegraphics[width=5.5cm]{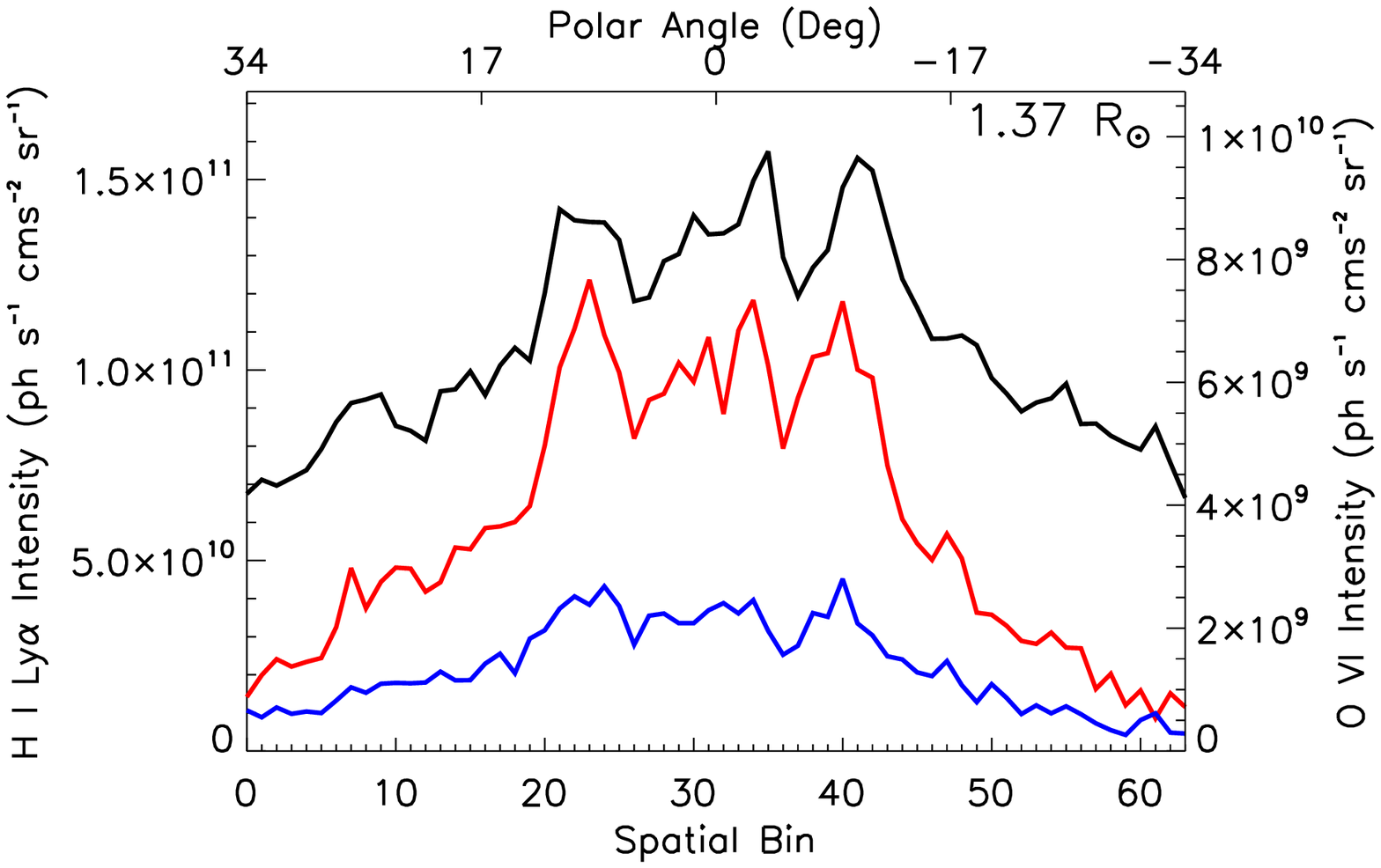}
  \includegraphics[width=5.5cm]{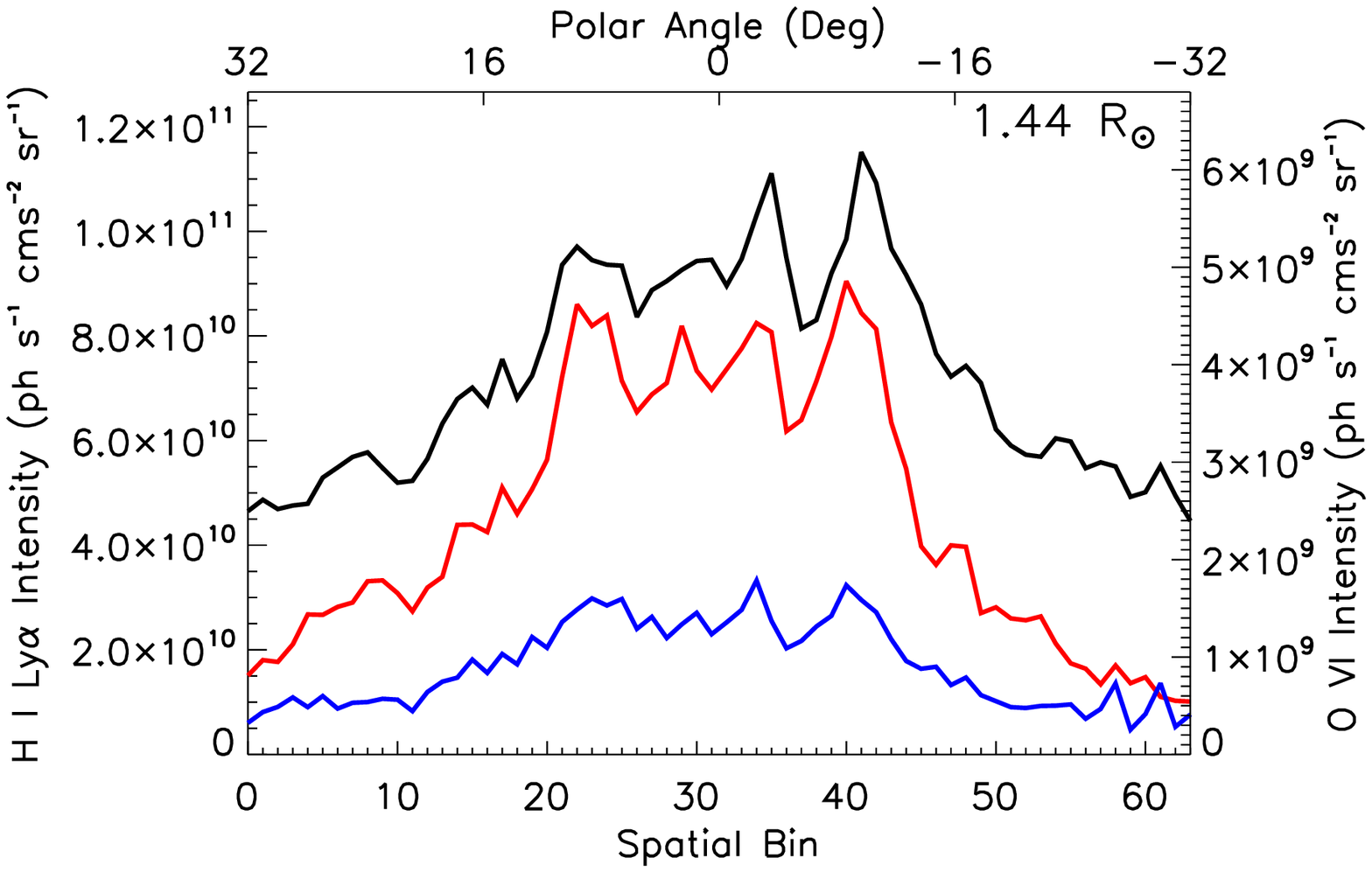}
  \includegraphics[width=5.5cm]{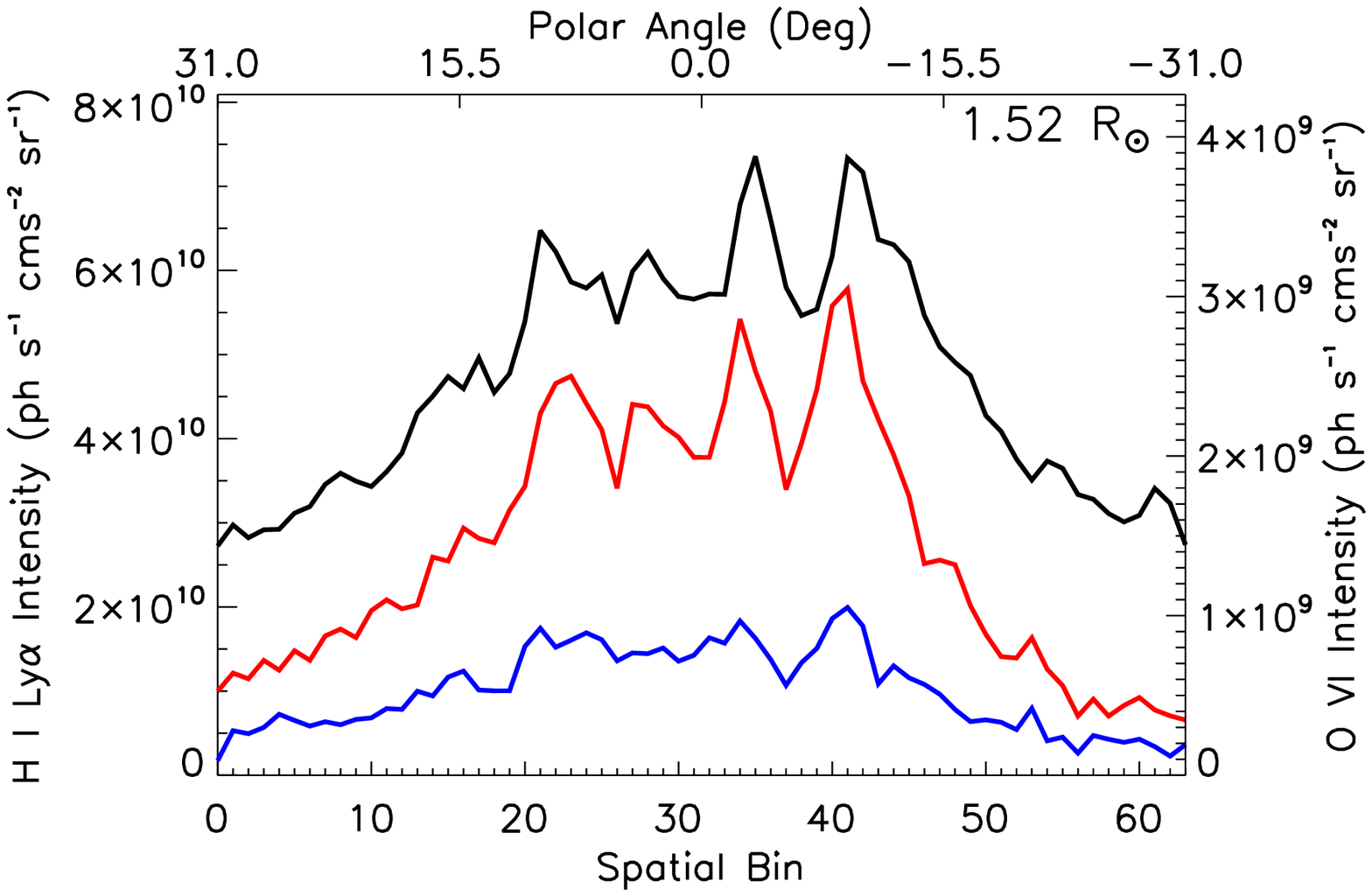}

  \includegraphics[width=5.5cm]{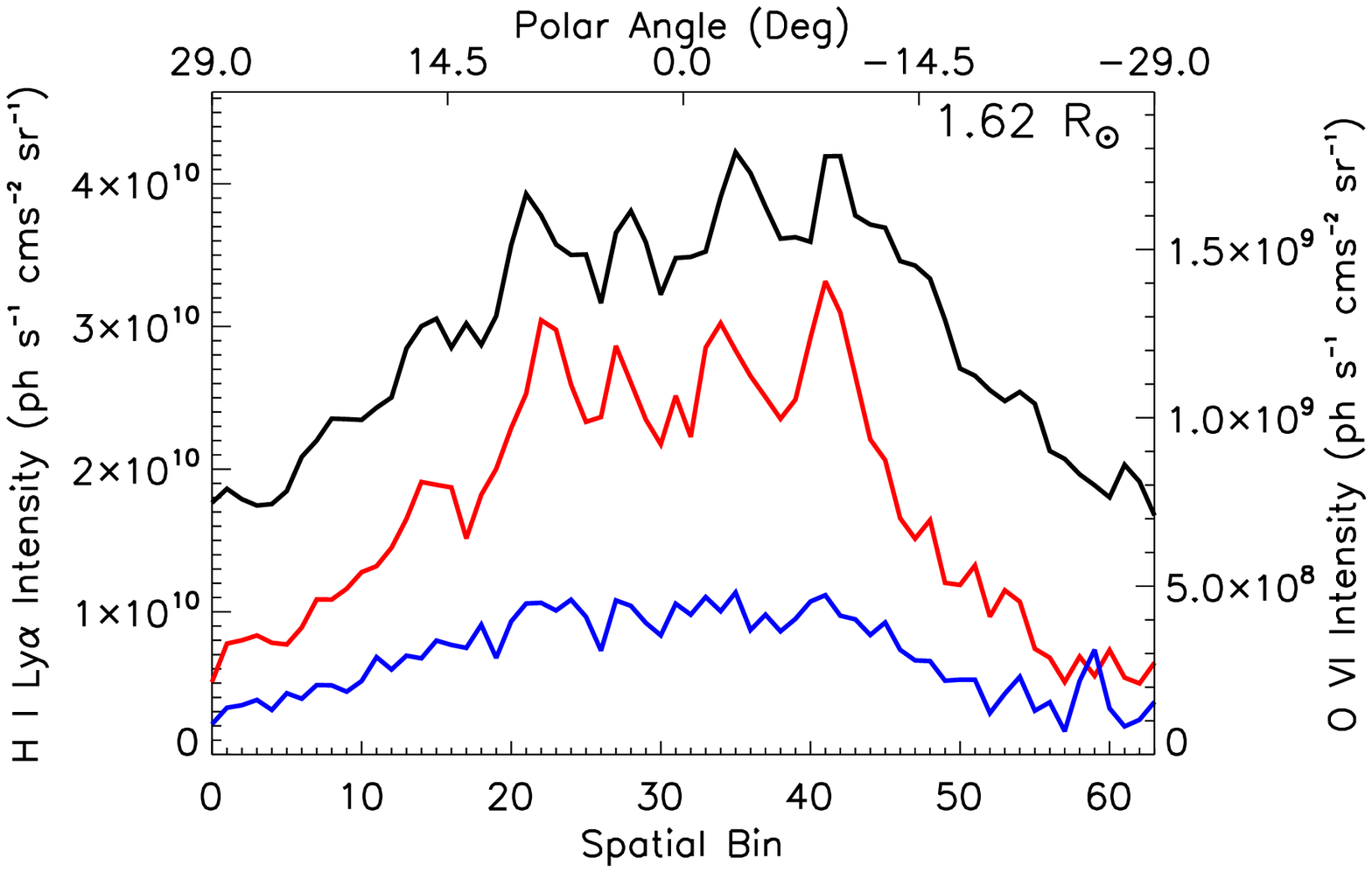}
  \includegraphics[width=5.5cm]{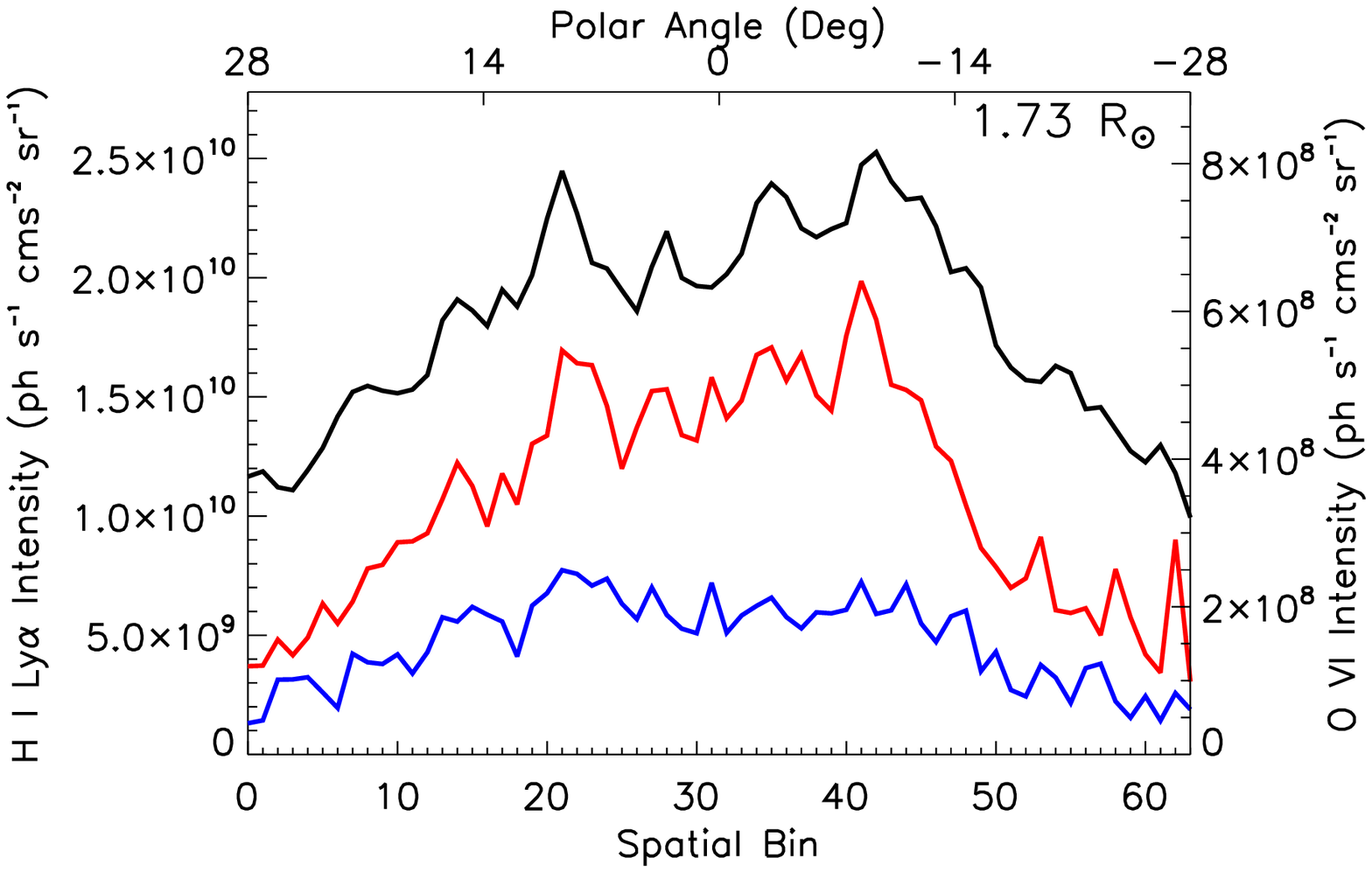}
  \includegraphics[width=5.5cm]{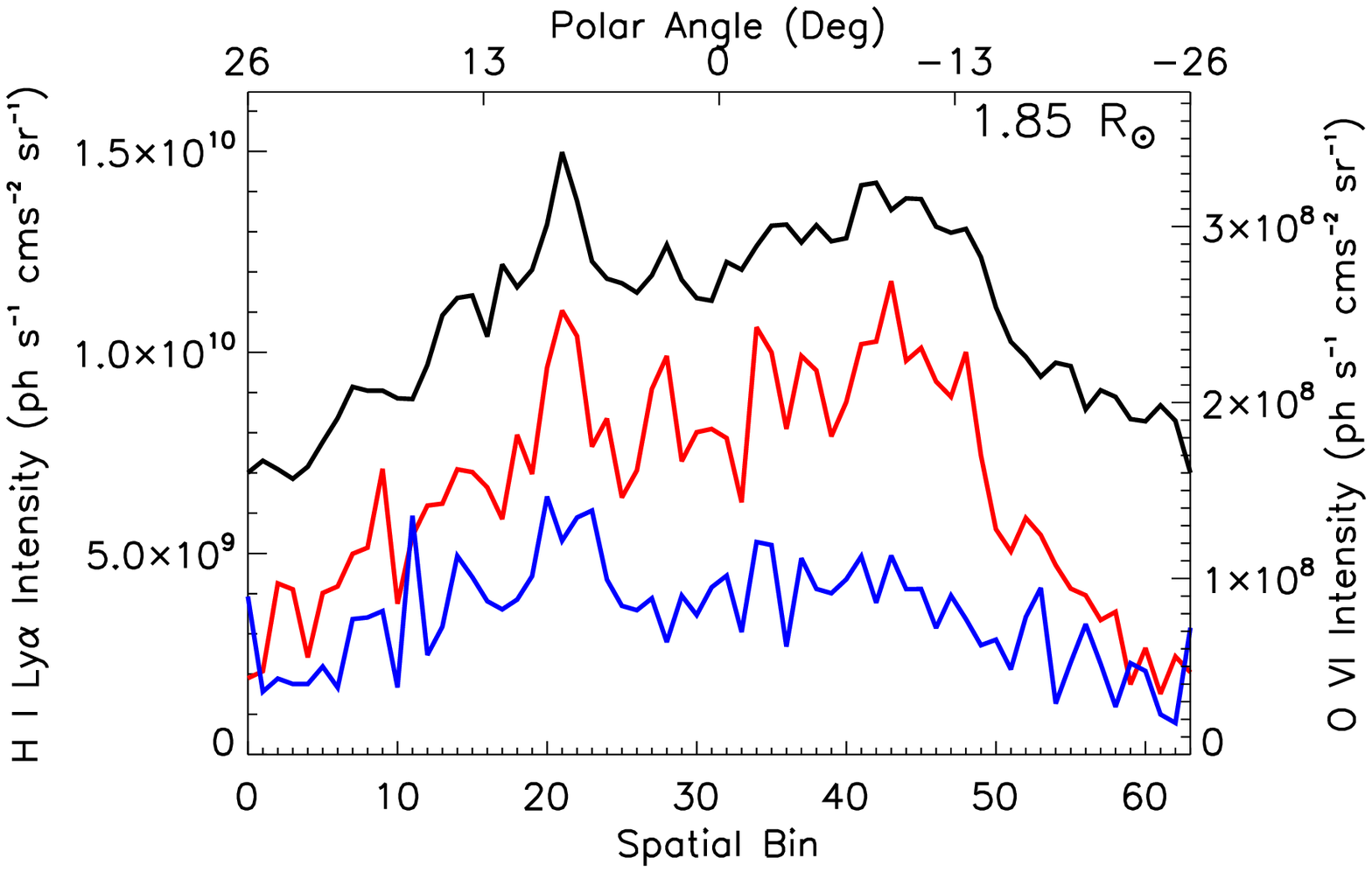}

  \caption{Intensity profiles in the north polar coronal hole, along the UVCS slit for six heliocentric distances. Intensity profiles of the redundant H~{\sc{i}}~Ly$\alpha$ are shown as black solid lines. The O~{\sc{vi}} 1031.9 and 1037.6 doublet intensities are represented by red and blue solid lines, respectively. We limited the sample of our data to the observations performed on the 6 and 7 April, with the slit center covering heliocentric distances up to 1.85~\rsun. The abscissa axis reports the polar angle latitude in degrees (upper axis) measured counterclockwise from the north pole, and the corresponding detector bins (bottom axis).}
  \label{fig:uvcs_data}
\end{figure*}

The intensity profiles along the slit after correction for the detector bias are shown in Fig. \ref{fig:uvcs_data} in the  O~{\sc{vi}} 1031.9, 1037.6~\AA, and H~{\sc{i}}~Ly$\alpha$ 1215.7~\AA\, lines at six heliocentric distances.

\subsection{SoHO LASCO-C2 and EIT context images}
\label{sec:context}

We used observations obtained by the SoHO instruments LASCO (Large Angle and Spectrometric COronagraph) C2 coronagraph (\citealt{Brueckner1995a}), and EIT 195\AA\ (Extreme ultraviolet Imaging Telescope; \citealt{Delaboudiniere1995a}) as context data, helping to discriminate the counterparts of plumes and interplumes in the UVCS spectra.
The three instruments have different fields of view, hence a single plume or interplume must be identified at different coronal altitudes, where it is imaged at different latitudes. Our starting point for plume identification is the occurrence of elongated bright emission regions in the polar caps in LASCO/C2 images. We assumed that plumes follow the profiles of the local magnetic field lines emerging from the coronal hole, and identified plumes across the field of view of the three instruments by tracing back model magnetic field lines from LASCO/C2 to the UVCS and EIT fields of view (see Section \ref{sec:ident}).

\subsection{Identification of plumes and interplumes}
\label{sec:ident}

The UVCS intensity profiles of Figure \ref{fig:uvcs_data} show a system of four candidate plumes and interplume regions, which can be identified both in O~{\sc{vi}} 1031.9~\AA\, and H~{\sc{i}}~Ly$\alpha$ 1215.7 \AA\, over an angle of roughly $\pm 14$ deg of amplitude, starting from the radial through the  north pole. The positions of these candidates, in the field of view of UVCS, are listed in Table~\ref{tab:02}, and the four plumes are labeled PL0, PL1, PL2, and PL3, and the interplumes IP0, IP1, IP2, and IP3. Radial and latitudinal positions of each identified structure are given at seven different altitudes, $h$, of the UVCS slit center in the corona.

We adopted the magnetic field model proposed by \cite{Banaszkiewicz1998a} to reproduce the plume profiles across the field of view of LASCO, UVCS, and EIT. This model consists of an axisymmetric combination of a dipole plus a quadrupole, with an azimuthal current sheet in the equatorial plane. Different geometries can be explored by changing the parameter $Q$ (see Eqs. 1 and 2 in \citealt{Banaszkiewicz1998a}), which controls the topology of the magnetic field lines, describing a pure dipole plus a current sheet when $Q=0$, and a combination of dipole plus a current sheet plus a quadrupole, with the contribution of the latter component increasing with $Q$, when $Q>0$.

\begin{figure*}[t]
  \includegraphics[width=8.5cm]{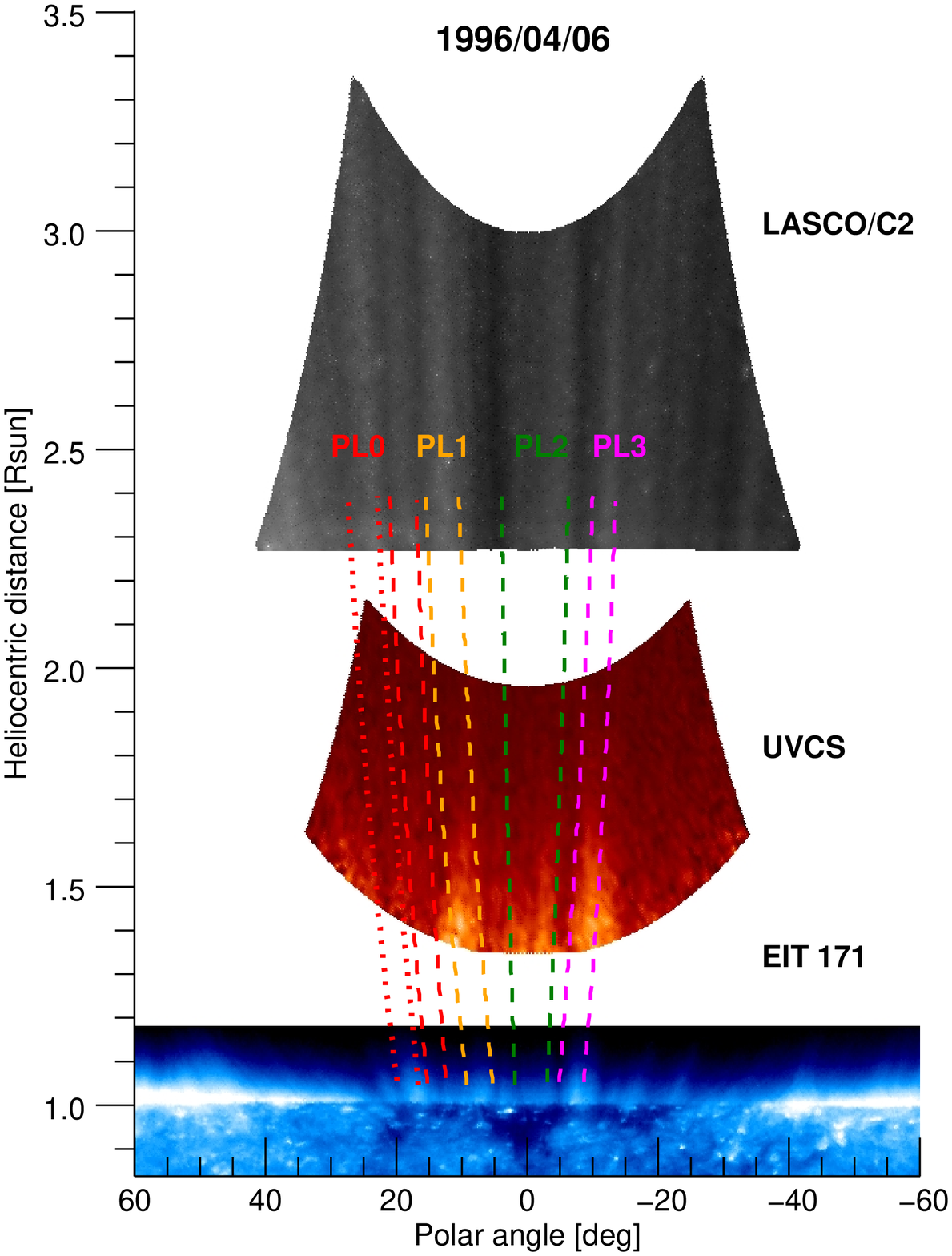}
  \includegraphics[width=8.5cm]{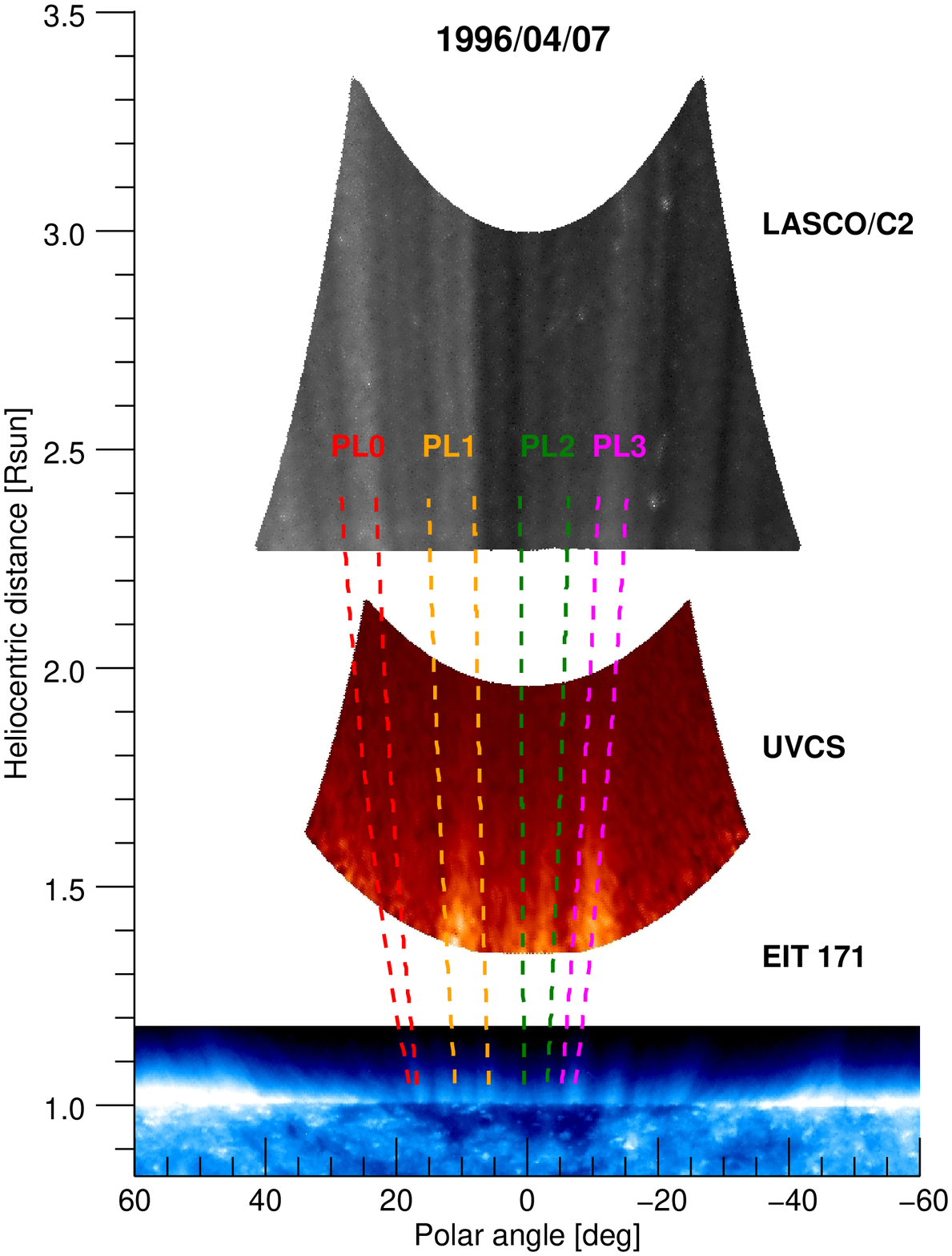}
  \caption{Composite UVCS, LASCO/C2, and EIT 195 \AA\ images of the solar corona, taken on 6 and 7 April, left and right panel, respectively. The plume identification has been made easier by stretching the contrast in the UVCS, LASCO, and EIT images, and by mapping them in polar coordinates, as the spatial scales of the three instruments are different. We point out that the UVCS image in the O~{\sc{vi}}~1031.9 \AA\ line has been obtained by juxtaposing the intensity profiles along the slit, as the covered coronal fields of the observations turned out to be at contiguous altitudes, and not by interpolation. The fits of the magnetic field lines of the \cite{Banaszkiewicz1998a} model to the plumes we identified in LASCO, UVCS and EIT images (see Table \ref{tab:02}) are shown as dashed lines. In the left panel, plume PL0 appears to split into two components in the LASCO and EIT images, and the fitted line profiles of the east side component has been shown in the panel as a dotted line. In UVCS data the PL0 east side component has not been considered owing to the low signal.
  }
  \label{fig:eit_uvcs_lasco}
\end{figure*}

The fit of the magnetic field lines to the plume profiles has been obtained by minimizing the mean of the distances, on the plane of the sky, of a sample of coordinate points along the observed plume profiles derived from the three instruments above, to a model of a magnetic field line, obtained by varying the latitude of the footpoint and the $Q$ parameter. The results of the fit are summarized in Table \ref{tab:03}. The typical values of the mean of the distances of the modeled field lines to the plume profiles turned out to be on the order of $24^{\prime\prime}$ on the plane of the sky. We obtained values for the $Q$ parameter and hence for the magnetic field geometry, ranging from a configuration characterized by a pure dipole plus a current sheet to configurations where in addition a contribution from a quadrupole is present. The circumstance of having obtained a set of different values for $Q$, together with the low number of analyzed plumes, prevents us from inferring precise information about the real magnetic field geometry in the coronal hole, based on the model by \cite{Banaszkiewicz1998a}. We hypothesize that the dispersion of the fitted values of $Q$ could be ascribed to a projection effect of the profiles on the plane of the sky, as different longitude angles of the plume footpoints imply a different projection on the plane of the sky of profiles that are otherwise similar. The fit to the plume profiles applied to LASCO/C2, UVCS, and EIT data is illustrated in Figures \ref{fig:eit_uvcs_lasco} a and b, showing two composite images of the three instruments, taken on 6 and 7 April 1996.

The positions of plumes appear to be stable in time and they can be recognized at the same positions in latitude during the two days of observations. Plume PL0 is an exception as it tends to move away from the pole on the 7 April and to split into two parts. From the EIT image, we can clearly identify PL0, PL1, and PL3 in the low corona, and their positions are
reasonably well traced back by the model magnetic field lines, starting from the LASCO images. During the 7 April, PL2 shows up as a faint structure, but it is still clearly identifiable in LASCO images. On the contrary, PL3 enforces and widens on the 7 April. Interplume lanes are well defined during the two days of observations. Interplume IP2 starts being visible from slit position 3 onward, perhaps because of the appearance of a background plume in between IP1.

\begin{table*}
\centering
\caption{Identification of plumes and interplumes, labeled as PL and IP,
  respectively, at different heliocentric distances, as observed in redundant
  Lyman $\alpha$ line. For each region we report its extension in spatial bins
  along the UVCS slit.
  }
\begin{tabular}{c c c c c c c c c c}
\hline\hline
slit &   h  & PL0        & PL1         & PL2          & PL3        & IP0    & IP1  & IP2    & IP3    \\
\hline
\hline
 0  & 1.375 & 15-18 & 21-25 & 30-35 & 38-43 & -              & 26-28 & -             & 36-37 \\
 1  & 1.442 & 14-17 & 21-25 & 29-35 & 39-44 & 18-19 & 26-28 & -             & 36-38 \\
 2  & 1.524 & 14-17 & 20-24 & 30-36 & 40-46 & 18-19 & 26-27 & -             & 37-38 \\
 3  & 1.620 & 13-16 & 19-24 & 31-37 & 40-46 & 17-18 & 26-27 & 29-30 & 38-39 \\
 4  & 1.730 &  4-10 & 17-23 & 34-37 & 41-46 & 12-14 & 26-31 & -             & 38-39 \\
 5  & 1.855 &  2- 9 & 17-23 & 32-33 & 43-49 & 12-13 & 26-30 & -             & 39-40 \\
\hline
\end{tabular}
\label{tab:02}
\end{table*}

The effect of solar rotation on the plume orientation can be considered negligible, particularly in the case of data acquired up to 2~\rsun, covering a time interval of 1.4 days from the first to the last exposure. However, we know that polar plumes tend to appear in the same location, with a typical life time of about one day (\citealt{Lamy_P_1997a}, \citealt{DeForest2001a}); hence, we do not know, when comparing observations of two successive days, if the plume that we see as stable at a given position is actually the same or a recurrent plume.

From the Table \ref{tab:02} data, we can estimate the typical size of the observed plumes projected on the plane of the sky, in the field of view of UVCS. At the coronal altitude of 1.4~\rsun, plume widths span from four to seven spatial bins along the slit, corresponding to a linear size perpendicular to the plume axis between $5\times 10^9~{\rm cm}$ and $1\times 10^{10}~{\rm cm}$ (the spatial bin is $28\arcsec$ wide along the slit).

\begin{table}
  \caption{Footpoint latitudes and $Q$ parameter obtained by fitting the magnetic field model of \cite{Banaszkiewicz1998a} to the observed plume profiles from LASCO/C2, UVCS, and EIT data. The footpoint latitudes are given in degrees of polar angle. The values are given for the east and west sides of the four plumes we studied, and are relative to the 6 and 7 April 1996.}
\centering
\begin{tabular}{l r r r r}
\hline\hline
6 April, 1996 & ~ & ~ & ~ & ~\\
\hline
plume               & ${\rm Lat_{east}}$ & ${\rm Q_{east}}$ & ${\rm Lat_{west}}$ & ${\rm Q_{west}}$ \\
 ~ & (deg) & ~ & (deg) & ~ \\
\hline
PL0 &   15.0 &  0.0 &   12.1 &  0.0 \\
PL1 &    8.9 &  0.5 &    5.1 &  1.2 \\
PL2 &    1.8 &  1.5 &   -2.9 &  1.5 \\
PL3 &   -4.6 &  1.5 &   -8.3 &  0.3 \\
\hline
7 April, 1996 & ~ & ~ & ~ \\
\hline
plume               & ${\rm Lat_{east}}$ & ${\rm Q_{east}}$ & ${\rm Lat_{west}}$ & ${\rm Q_{west}}$ \\
~ & (deg) & ~ & (deg) & ~ \\
\hline
PL0 &   17.4 &  0.3 &   16.5 &  0.0 \\
PL1 &   10.8 &  0.0 &    5.8 &  0.0 \\
PL2 &    0.5 &  1.5 &   -2.9 &  1.5 \\
PL3 &   -5.0 &  1.5 &   -7.0 &  1.5 \\
\hline
\end{tabular}
\label{tab:03}
\end{table}

The expansion of plumes with altitude in the corona can be described by means of their expansion factor, which is defined as

\begin{equation}
  f(r,\theta)=\frac{A(r,\theta_0)}{A(R_\odot,\theta_0)}
\left({
\frac{R_\odot}{r}}\right)^2
,\end{equation}

\noindent where $A(r,\theta)$ is the cross section of the plume at distance $r$ from the Sun, and at latitude $\theta$. The latitude $\theta_0$ is that of the footpoint of the field line on the solar surface. From the requirement of the magnetic flux conservation, $B(r,\theta)A(r,\theta)=B(R_{sun},\theta_0)A(R_{\rm sun},\theta_0)$, we have

\begin{equation}
  f(r,\theta)=
  \frac{B(R_\odot,\theta_0)}
  {B(r,\theta)}
  \left({
    \frac{R_\odot}{r}
  }\right)^2
  \label{eq:expansion}
,\end{equation}

\noindent where $B(r,\theta)$ is, in our case, the intensity of the magnetic field given by the model of \cite{Banaszkiewicz1998a}. An estimate of the expansion factor of plumes has been done, for example, by \cite{DeForest1997a}, who derived $f\sim 3$ at 4-5~\rsun. The expansion factor predicted by the magnetic field model we adopted, at the heliocentric distance of 5~\rsun, is $f=2.7$ for $Q=0$, hence in good agreement with the observations by \cite{DeForest1997a}; however, when $Q=1.5$ the predicted value is $f=8.1$.

We can make a comparison of the typical size of plumes we obtained from UVCS data with observations in the low corona (see, e.g., \citealt{Wilhelm2011a}, \citealt{Poletto2015a} and references therein), which suggest a cross section of plumes of $\sim 30$~Mm. If we extrapolate this value by means of our model up to 1.4~\rsun, we obtain $5\times10^9~{\rm cm}$, in agreement with the cross section of the  plumes as seen by UVCS at this altitude.

\subsection{Filling factor of the plumes}

\begin{figure}[t]
  \centering
  \includegraphics[width=8.0cm]{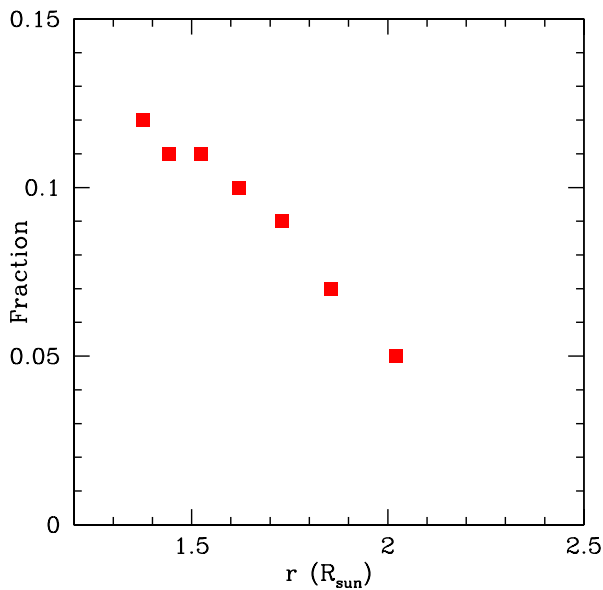}
  \caption{Fraction of coronal hole cross section covered by plumes, at
    different UVCS slit altitudes.}
  \label{fig:filling}
\end{figure}

The filling factor of the plumes in the coronal hole we are studying, which is the fraction of the coronal hole volume occupied by plumes, can be estimated by making the assumptions that we are observing all the plumes present in the coronal hole, projected on the plane of the sky, and that they can be described as a cylindrical shaped structures, as the only information we have is their appearance on the plane of the sky. We used the data contained in Table~\ref{tab:02} to evaluate the cross section of each plume and to calculate the total volume occupied by plumes, compared to the volume of the portion of the coronal hole we considered. We derived filling factors at different heliocentric distances, and the resulting values are shown in Figure~\ref{fig:filling}. The plume filling factors range from 0.12 at 1.375~\rsun, to 0.07 at 2.194~\rsun, decreasing with distance, and they are consistent with a commonly accepted estimate of 0.1 (see \citealt{Wilhelm2011a}).

A weakness of this method of evaluating the filling factor is partly related to the fact that, owing to the integration along the line of sight, we do not know how many background or foreground plumes are hidden or not resolved. In this regard, we add that, on the basis of SoHO/EIT observations, \cite{Gabriel2009a} proposed the existence of two distinct classes of plumes: beam plumes, that have a cylindrical structure, and network plumes, that are a curtain shape and become visible when they are aligned along the LOS, and whose extension can be underestimated. We point out that the analysis made by \cite{Gabriel2009a} refers to altitudes in the corona in the range of $1.02-1.08 ~{\rm R_\odot}$, hence we do not know if this distinction is relevant for our study. Another difficulty arises with the definition of the filling factor concept itself, as
distinguishing coronal plasma in plumes and interplumes could arbitrarily simplify a more complex scenario of coronal structures not resolved in space and time, as demonstrated by \cite{Feldman_1997a}. However, counting visible plumes and measuring their size is a relatively straightforward approach, providing at least a lower limit to the contribution of plumes to the polar wind.

\subsection{Intensity ratio of O~{\sc{vi}} doublet}
\label{sec:doublet}

The intensity ratio of the O~{\sc{vi}} doublet, $R=I_{1032}/I_{1038}$, can be used to discriminate between low and high outflow speed regimes in the coronal plasma. A comprehensive description of the use of $R$ as a diagnostic tool can be found in \cite{Noci1987a}. Owing to the dimming of the radiative component of the O~{\sc{vi}} doublet lines, the $R$ values decrease with the outflow speed and we can broadly state that values of about $R=3-4$ correspond to an almost static corona, while $R=2$ is attained for an outflow speed of about $100~{\rm km/s}$. Short descriptions of the formation of UV lines  in the corona and
of the Doppler Dimming effect are given in Sects. \ref{sec:l_form} and \ref{sec:dd_analysis}, respectively. The intensity ratio, $R$, for our purposes can be considered to not be dependent on the choice of a particular coronal model, as it is only slightly affected by uncertainties in the chromospheric excitation radiation, in the electron density, and in the electron temperature.

\begin{figure*}
  \centering
  \includegraphics[width=20.0cm]{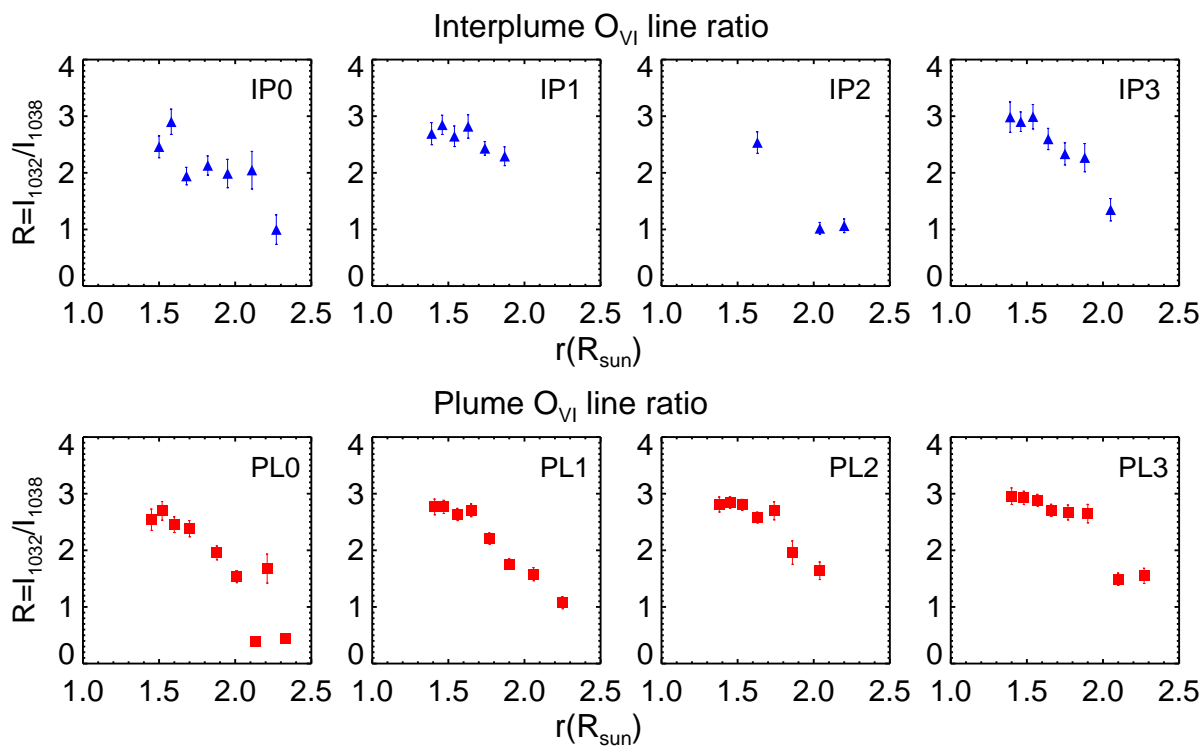}
  \caption{Intensity ratio of O VI doublet line for plumes and interplumes identified in Table (\ref{tab:02}). The heliocentric distances of data are calculated where a plume or a interplume meets the UVCS slit. The errors in the ratio are a function of the intensity of the corona and of the exposure time, which is larger for high heliocentric distances (see Table \ref{tab:01}).}
  \label{fig:ratio}
\end{figure*}

The profiles of $R$ in the plumes and interplumes we identified in UVCS data are shown in Figure \ref{fig:ratio}. The four interplumes are characterized by a linear decrease of $R$ with distance, pointing to a constant plasma acceleration with the altitude in the corona. The behavior of plumes is different as the profiles exhibit a less evident decrease with distance, which becomes more pronounced at coronal altitudes higher then 1.7-1.8~\rsun, suggesting that plumes are characterized by a low outflow speed regime in the interval of distances 1.5-1.8~\rsun, and after that are subjected to an increasing acceleration.

\subsection{Line profile analysis}
\label{sec:line_profiles}

\begin{figure*}
  \centering
  \includegraphics[width=16.0cm]{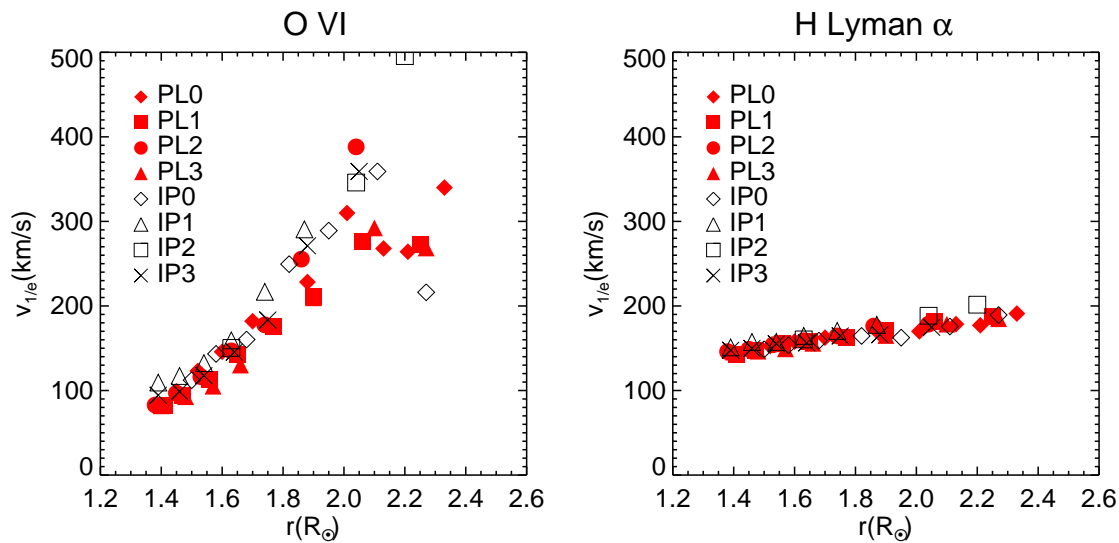}
  \caption{Lines widths of O VI and H Lyman alpha, given as microscopic speed of the ions at $1/e$ of the line profile height. The widths are corrected for the instrumental function and for the slit width.}
  \label{fig:most_prob_speed}
\end{figure*}

We performed an analysis of the line profiles by fitting observed spectra with a Voigt function, which comprehends the instrumental function, the slit contribution, and the intrinsic line width (see \citealt{tesi_phd_silvio}). The intrinsic line widths of the O~{\sc{vi}}~1031.9 \AA\ and H~{\sc{i}}~Ly$\alpha$ lines are shown as most probable speeds in Figures~(\ref{fig:most_prob_speed}a,b).  In oxygen ion observations, we can see a clear distinction between plume and interplume line widths, with the plumes attaining lower values on average, corresponding to lower kinetic temperatures in plumes than interplumes, in agreement with \cite{Giordano2000a}, who found ion speed distribution broader in interplumes than in plumes. An exception is represented by the data on plume PL1, which shows a radial profile similar to that of interplumes, for distances higher 1.6~\rsun. The difference between plumes and interplumes is less evident in H~{\sc{i}}~Ly$\alpha$, and line profiles appear almost indistinguishable within errors (see Figure \ref{fig:most_prob_speed}b). The results are in agreement with what was expected on the basis of previous studies (see, e.g., \citealt{Wilhelm2011a}, \citealt{Poletto2015a}). In particular, \cite{Hassler1997a} have demonstrated that for heliocentric distances higher than 1.1-1.3~\rsun~ plumes may have kinetic temperatures lower than interplumes (see also \citealt{Kohl1997a}) and, on the contrary, close to the coronal base the plume temperatures are higher than those of interplumes (see also \citealt{Walker_1993a}).

\section{Data analysis}
\label{sec:analysis}

\subsection{Formation of emission lines in the intermediate corona}
\label{sec:l_form}

The coronal O~{\sc{vi}} doublet and H~{\sc{i}}~Ly$\alpha$ lines form by the collisional excitation of ions and by resonance scattering of the radiation coming from the chromosphere. The emissivity of the collisional component, $j_{\rm coll}$, is a function of the electron density, $n_{\rm e}$, the electron temperature, $T_{\rm e}$, through the ionized fraction $R_{\rm ion}(T_{\rm e})=n(X)/n(X^{+m})$, and the element abundance relative to hydrogen $A_{\rm el}=n(X)/n_{\rm H}$. The emissivity of the resonantly scattered component, $j_{\rm res}$, also depends on the outflow speed of the plasma element in the corona, $w$, and on the absorption profile of coronal ions, which is determined by the local kinetic ion temperature, $T_{\rm ion}$. The outflow speed is responsible for a reduction of the resonant scattering efficiency. Hence, the radiative component of a coronal line can be used as a diagnostic tool, called Doppler dimming (DD) analysis, to infer $w$ (see, e.g., \citealt{Hyder1970a}, \citealt{Noci1987a}).

In the case of H~{\sc{i}}~Ly$\alpha$, resonant excitation accounts for 99\% of the total intensity (see \citealt{Gabriel1971a}). A detailed account of the formation mechanism of these lines can be found in \cite{Withbroe1982a}, \cite{Kohl1982a}, \cite{Noci1987a}, and \cite{Noci1999a}.
The total line intensity, $I_{\rm line}$, is the sum of the collisional and radiative contributions integrated over the frequency and along the line of sight:

\begin{eqnarray}
\nonumber
  I_{\rm line}(n_{\rm e},w,A_{\rm el},T_{\rm e},T_{\rm ion})&=&I_{\rm coll}+I_{\rm res}\\
  &=&\int_{\rm LOS}\int_\nu
  \left({
    j_{\rm coll}+j_{\rm res}
  }\right)
  d\nu dx
  \label{eq:tot}
.\end{eqnarray}

\subsection{Doppler Dimming analysis}
\label{sec:dd_analysis}

We inferred the radial outflow speed and the electron density of the plasma in plumes and interplumes by means of the DD analysis applied to the O~{\sc{vi}} doublet and H~{\sc{i}}~Ly $\alpha$ lines. We make use of a method for separating the collisional and radiative components of the observed intensities, determining self-consistently the electron densities and the outflow speeds of the coronal plasma. In this section we will only give a summary of the method,  a detailed description of which can be found in \cite{Zangrilli2002a}.

The separation of the radiative and collisional components can be obtained from the inversion of Eq.~(\ref{eq:tot}), by writing the O~{\sc{vi}} doublet intensities $I_{1031.9}$ and $I_{1037.6}$, as

\begin{eqnarray}
\nonumber
I_{1031.9}
& = &
2{\cal C}(n_{\rm e})+4{\cal R}(n_{\rm e},w_{\rm OVI})\\
I_{1037.6}
& = &
{\cal C}(n_{\rm e})+{\cal R}(n_{\rm e},w_{\rm OVI})+{\cal P}(n_{\rm e},
w_{\rm OVI})
\label{sys:int}
,\end{eqnarray}

where ${\cal C}(n_{\rm e})$ is the intensity of the collisional component of the O~{\sc{vi}}~$\lambda$1037.6 line, and  ${\cal R}(n_{\rm e},w_{\rm OVI})$ and ${\cal P}(n_{\rm e},w_{\rm OVI})$ are the radiative components of the O~{\sc{vi}}~$\lambda$1037.6 line, excited by the corresponding chromospheric O~{\sc{vi}} line and by the C~{\sc{ii}} chromospheric pumping lines, respectively. We note that ${\cal C}$ is a function of the electron density squared only, while ${\cal R}$ and ${\cal P}$ depend linearly on $n_{\rm e}$ and also on the outflow speed, $w$. The components ${\cal C}$, ${\cal R,}$ and ${\cal P}$ are uniquely determined by a couple of values $\left({n_{\rm e},w_{\rm OVI}}\right)$, if all other model parameters are known, hence the solution of the system of equations (\ref{sys:int}) is uniquely determined by the two observed quantities, $I_{1031.9}$ and $I_{1037.6}$. \citealt{Zangrilli2002a} provide a discussion on the existence and the uniqueness of the solutions.
Assuming that the H~{\sc{i}}~Ly$\alpha$ line is essentially formed by resonant scattering of the chromospheric radiation, given the $n_{\rm e}$ derived from the DD applied to the O~{\sc{vi}} doublet, we can calculate the proton outflow
speed.
The search for a solution of equations (\ref{sys:int}) starts with a first guess coronal model, then the predicted intensities ${\cal C}$, ${\cal R,}$ and ${\cal P}$ are used to derive new values for $n_e $ and $w$; the procedure is iterated until the observed intensities $I_{1031.9}$ and $I_{1037.6}$ are reproduced.

We briefly recall the essential physical parameters of the coronal model needed for the DD analysis. The intensities of the UV coronal lines depend primarily on the plasma density and the outflow speed, which we find by means of the DD analysis, and the element abundance, which we determine by the constraint on the mass flux conservation (see Section \ref{sec:mass_flux}). A somewhat secondary role is played by the plasma electron temperature. We assume the electron temperature radial profile according to \cite{Cranmer1999a}, which is typical of a polar coronal hole. A further assumption is that the oxygen ion populations in coronal holes freeze in at the electron temperature of $T_{\rm e}=1.1~{\rm MK}$, as has been demonstrated by \cite{von_Steiger2000a} (see also \citealt{Ko1999b}). In our model, the electron temperature value $T_{\rm e}=1.1~{\rm MK}$ is attained at 2.35\rsun, hence for higher heliocentric distances the O~{\sc{vi}} abundance is frozen in. The neutral hydrogen is not supposed to be frozen.

A summary of the plume and interplume measurements of $T_{\rm e}$ is given by \cite{Wilhelm2011a} (see Table 4 therein) for heliocentric distances up to 1.13\rsun, from which we deduce that typical electron temperatures of plumes are about 0.8 MK, almost constant with height, while those of interplumes tend to increase with distance and to be higher with respect to plumes, from 10 to 30\%. Unfortunately, we have no information at altitudes higher than 1.13\rsun. Given the uncertainties in the determinations of $T_{\rm e}$ and the lack of information in the interval of distances covered by UVCS observations, we chose not to distinguish between plume and interplume electron temperature, and the radial profile we adopted is representative of the mean conditions in a polar coronal hole; we will discuss the impact of this choice on the results. The ion kinetic temperature radial profiles included in the model, are obtained by fitting the most probable speed line profiles, as discussed in Section (\ref{sec:line_profiles}).

The value of the solar chromospheric intensity of H~{\sc{i}}~Lyman~$\alpha$ has been taken from the UARS/SOLSTICE (Upper Atmosphere Research Satellite/SOLar STellar Irradiance Comparison Experiment) experiment (see \citealt{Woods2000a}). In the case of O~{\sc{vi}} doublet lines there are no routine measurements, hence we used the values of 387.0 and $199.5~{(\rm erg/s/cm^2/sr)}$, measured from UVCS disk observations taken on December 4, 1996, for the 1031.9 and 1037.6~\AA\ lines, respectively, (see \citealt{Zangrilli2002a}). The adopted chromospheric line widths are taken from \cite{Warren1997a}.

The amount of ion temperature anisotropy of the ion kinetic temperatures along and across the magnetic field direction in the intermediate corona has been debated in the literature, since the beginning of the SoHO mission. The profile widths of oxygen lines observed by UVCS, larger than previously expected, have been interpreted as strong evidence for preferential ion heating perpendicularly to the magnetic field lines direction (see, e.g., \citealt{Kohl1997a}). A different explanation for the observed large profiles has been proposed by \cite{Raouafi2004b} and \cite{Raouafi2006a}. The objections raised by \cite{Raouafi2006a} have been discussed by \cite{Cranmer2008a}, who concluded that the most probable temperature anisotropy ratios correspond to $T_{i\parallel}/T_{i\perp}\approx 6$, thus we adopt this value in our coronal model.

\subsection{Mass flux conservation}
\label{sec:mass_flux}

From the DD analysis (see Section \ref{sec:dd_analysis}) we derived the plasma outflow speeds and electron densities of plumes, at different heliocentric distances. These results should satisfy the constraint of mass-flux conservation, given a geometrical configuration of the polar plumes, for protons and oxygen ions separately. In general, the mass flux conservation for a generic atom in the corona can be written as
\begin{equation} 
  w(r,\theta)=\frac{F(1 AU,\theta)(215R_\odot)^2}{n(r,\theta)r^2}
  \frac{f(1AU,\theta)}{f(r,\theta)}
\label{eq:mass_cons}
,\end{equation}
where $w(r,\theta)$ and $n(r,\theta)$ are the atom outflow speed and density at distance $r$ and latitude $\theta$, respectively, $F(1 AU,\theta)$ is the atom mass flux measured at 1AU, which is assumed as a boundary condition, and $f(r,\theta)$ is the expansion factor, which describes the super-radial expansion of the plumes. A commonly accepted value for the fast wind proton flux is $F(1AU,\theta)\equiv F_p(1AU)=2\times 10^8~{\rm cm^{-2}~s^{-1}}$, although in the literature higher values are also given (see \citealt{Goldstein1996a}, \citealt{McComas2000a}), and we implicitly assume that it only depends on the radial coordinate $r$. The outflow speed of oxygen atoms at 1 AU turns out to be about 5\%-10\% faster than protons (e.g., Hefti et al. 1998). If we assume that, for our purposes, their speed is the same at 1 AU, from the condition of mass conservation (\ref{eq:mass_cons}) for oxygen and protons we have 
\begin{equation}
  w_{O^{+5}}(r,\theta)=w_p(r,\theta)\frac{Abb_{oxy}(1AU)}{Abb_{oxy}(r,\theta)}
\label{eq:oxygen_cons}
,\end{equation}
where $Abb_{oxy}=n_{oxy}/n_p$ is the abundance of oxygen with respect to hydrogen. Condition (\ref{eq:oxygen_cons}) does not depend upon knowledge of the expansion factor and ensures the mass flux conservation of oxygen atoms in the wind, provided that of protons is satisfied. The latter requirement can be verified after the DD analysis (see Section \ref{sec:results_geom}).

Knowledge of the geometry of plumes, described by the expansion factor $f(r,\theta)$, does not enter into the DD analysis, and the plume profiles derived from LASCO images, and traced back to the UVCS and EIT fields of view by means of the \cite{Banaszkiewicz1998a} model, have been used to identify the plumes in UVCS data (see Section \ref{sec:ident}).

\subsection{Distribution of plumes and interplumes along the line of sight}
\label{ssec:ipl_pl_models}

The intensities of UV spectral lines we derived from the UVCS observations are the sum of all coronal contributions along the LOS. When observing plumes, we can assume that we will measure the sum of different contributions coming from plumes and interplumes. We first derive an interplume model, based on the hypothesis that in the observed interplume regions the possible contribution of background plumes is negligible, and we identify the interplume plasma as the coronal hole medium surrounding plumes, assuming a spherical symmetry. Then we model plumes by confining the emitting plasma within a single cylindrical shaped structure located in the plane of the sky, embedded in the interplume medium.

\section{Results}
\label{sec:results}

\subsection{Plasma outflow speeds and electron densities}
\label{sec:results_dd}

\begin{figure}[t]
  \centering
  \includegraphics[width=8.0cm]{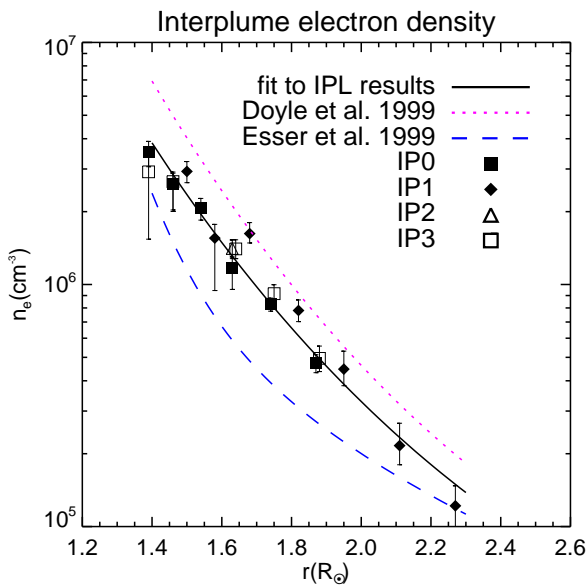}
  \caption{Electron density of interplume plasma from UVCS data. A comparison is made with $n_{\rm e}$ radial profiles taken from previous works.}
  \label{fig:ip_ne}
\end{figure}

The results obtained from the DD analysis described in Sect.~\ref{sec:analysis} are illustrated in Figs.~\ref{fig:ip_ne}, and \ref{fig:cons}. The main error source in the electron densities and outflow speeds we derived is the statistical uncertainty in the measured intensities of the O~{\sc{vi}} doublet lines, which can be estimated by means of the Poissonian statistics on the integrated counts at detector level. In the case of H~{\sc{i}}~Lyman~$\alpha$ the error given by the Poissonian statistics is negligible.

Further uncertainty is related to the radial profile of the electron temperature of our model. As far as the derivation of the O~{\sc{vi}} outflow speed by means of the DD technique is concerned, this mainly depends on the intensity ratio of the two O~{\sc{vi}} doublet lines, which is mostly independent of the electron temperature. However, the derived electron densities depend on $T_{\rm e}$, through the ionized fraction parameter $R_{\rm ion}(T_{\rm e})=n(X^{+m})/n(X)$, as the resonant component of the coronal UV line intensities is proportional to the product $R_{\rm ion}(T_{\rm e})A_{\rm el}n_{\rm e}$, and the collisional components on $R_{\rm ion}(T_{\rm e})A_{\rm el}n_{\rm e}^2$, where $A_{\rm el}$ is the element abundance. We can obtain a rough idea of the effect of our choice for $T_{\rm e}$, by assuming  $T_{\rm e}=0.8~{\rm MK}$ as a lower
limit for the electron temperature in plumes, which turns out to be a decrease in the electron density of about 30\%. This is because at those temperatures the ionization balance is a decreasing function, both for O\textsuperscript{+5} and for protons. For the same reason, the uncertainty in $T_{\rm e}$ does not much affect the calculation of the proton outflow speed, as a variation in the electron density is compensated for by an opposite variation in the ionized fraction.

The electron densities in interplumes derived from the DD analysis are shown in Figure~(\ref{fig:ip_ne}), and they are compared with the fitted interplume density profiles given by \cite{Doyle1999b} and \cite{Esser1999a}. With the purpose of establishing a model for interplumes, we performed the following power law fit of the electron densities we derived:

\begin{equation}
  n_{\rm e}(r)=1\times 10^5\times\left({\frac{435.98}{r^{7.31}}+\frac{2.50}{r^{2.22}}}\right)
  \label{eq:neipfit}
.\end{equation}

\begin{figure*}
  \centering  \includegraphics[width=16.0cm]{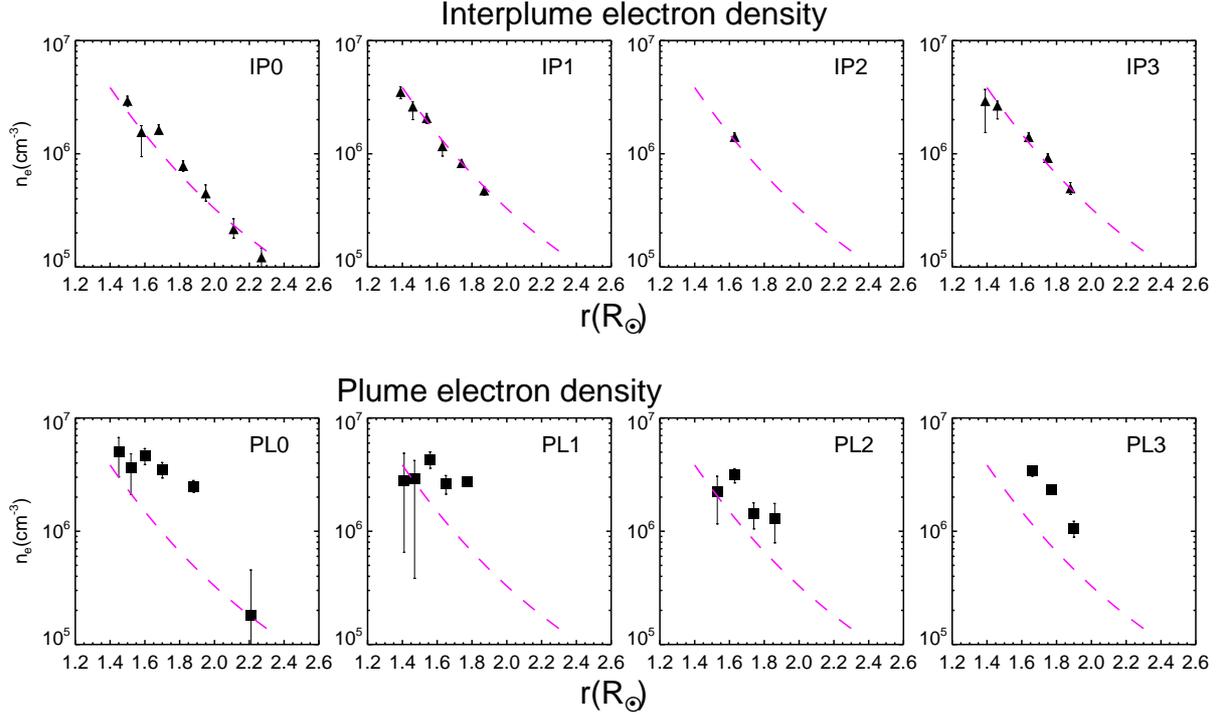}
  \caption{Electron density in interplumes (upper row) and plumes (lower row), for a model with a single cylindrical shaped plume in the plane of the sky. The red lines represent, for comparison, the interplume electron density profile, obtained by fitting all data with a power law (see Eq. \ref{eq:neipfit}).}
  \label{fig:density_multi}
\end{figure*}

\begin{figure*}
  \centering
  \includegraphics[width=16.0cm]{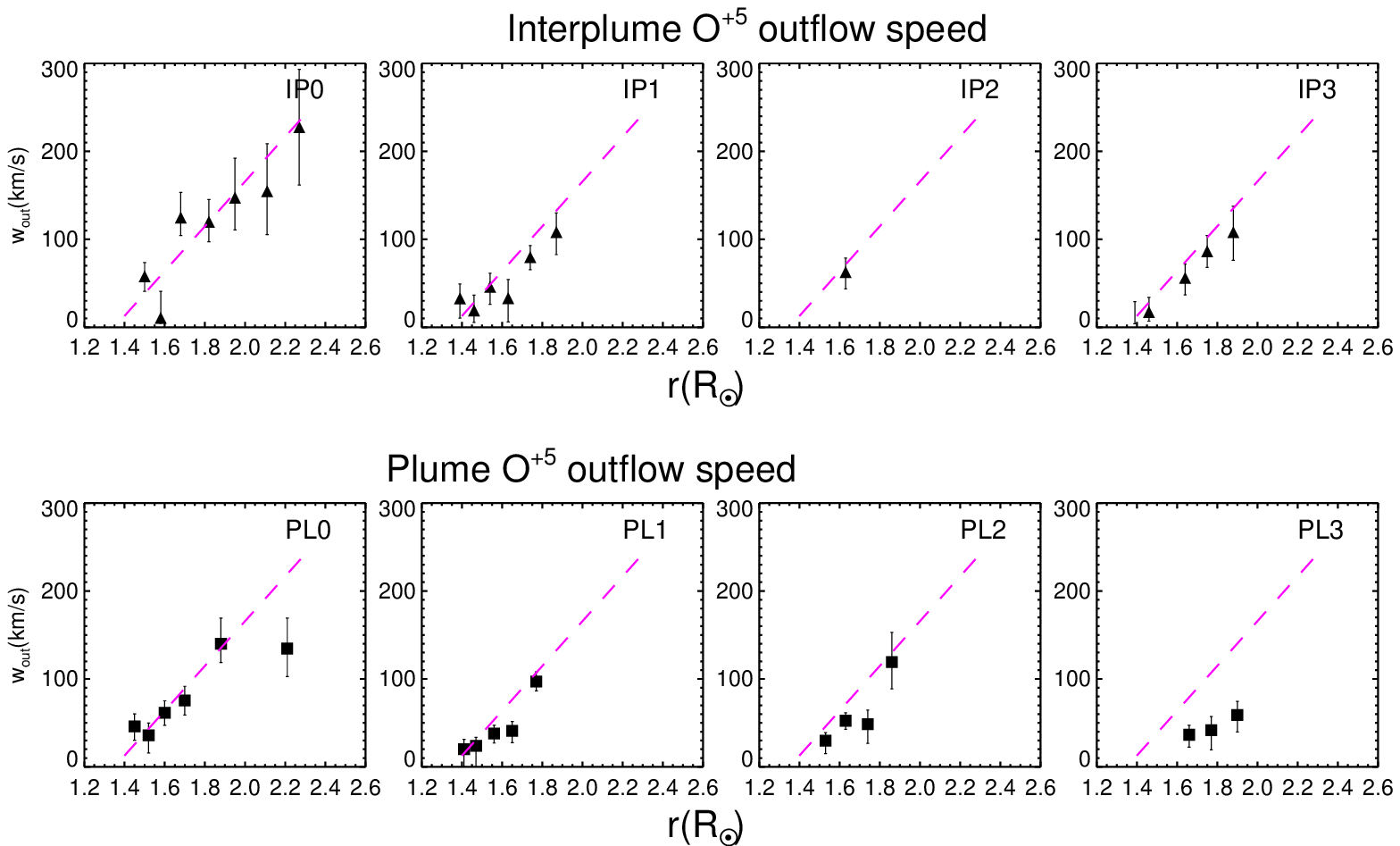}\\
  \includegraphics[width=16.0cm]{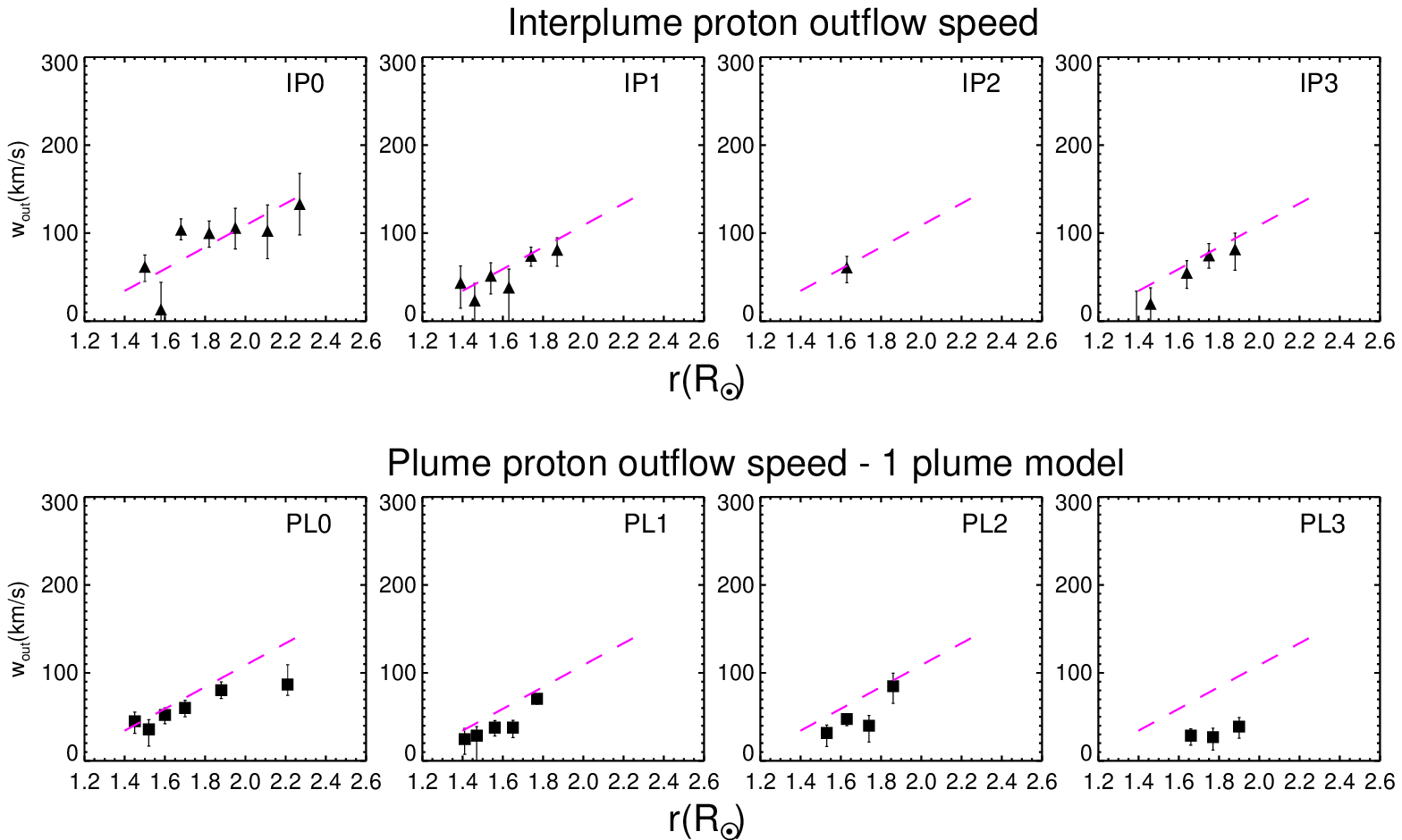}
  \caption{Outflow speed of O\textsuperscript{+5} in interplumes (first row from the top) and plumes (second row), in the case of a one plume model. The analogous results for protons are shown in the third and fourth rows. The red lines represent, for comparison, the interplume O\textsuperscript{+5} and protons outflow speed profiles (see Eqs. \ref{eq:voviipfit} and \ref{eq:vpipfit}, respectively).}
  \label{fig:ovi_speed_multi}
\end{figure*}

In the DD analysis of plumes, a combined plume-interplume emissivity along the LOS has been considered, where a single plume in the plane of the sky is embedded in the interplume medium (see Section \ref{ssec:ipl_pl_models}). The electron densities and outflow speeds for protons and O\textsuperscript{+5} ions are shown in Figs.~(\ref{fig:density_multi}), and (\ref{fig:ovi_speed_multi}), where in the first row the results for interplumes are represented. The second row shows the results for plumes. We immediately see that plumes are substantially denser than interplumes, and the density contrast turns out to be about 3-5, depending on the heliocentric distance, as expected on the basis of previous studies (see \citealt{Wilhelm2011a}, \citealt{Poletto2015a}).

As far as the outflow speeds are concerned, the basic result is that plumes have non-negligible outflow speeds, within the estimated errors, hence they can not be considered static structures, at least for distances higher than 1.6\rsun. Speed profiles for O~{\sc{vi}} ions of PL0 and PL1 are similar to the profiles typical of interplumes, while in the case of PL2 and PL3 they differ, showing a lower acceleration with distance. In general, both for plumes and interplumes the proton speed profiles attain lower values than O~{\sc{vi}} ions. The considerations that can be drawn, when comparing plumes and interplumes, are the same we made above for O~{\sc{vi}}.

We stress that we must consider the electron density values we derived from the DD analysis as mean values, as we know that plumes show time variability on timescales shorter than the integration times of our UVCS data, and space substructures on spatial scales well below the spatial resolution of UVCS (see Section \ref{sec:intro}). However, the outflow speeds of O~{\sc{vi}} ions, derived from the DD analysis, are not sensitive to the uncertainties in the absolute value of the electron density, as they mainly rely on the intensity ratio of doublet lines from the same ion. The case of neutral hydrogen is different, and the derived outflow speeds are dependent on the electron density. If the micro-scale filling factor within plumes is less then unity, that is, if substructures or inhomogeneities are present and they are not resolved by UVCS, we could underestimate the electron density, hence leading to an underestimate of the outflow speed of protons from the DD analysis. As has been stated, however, the derivation of the O~{\sc{vi}} ions outflow speed is not affected by the presence of a possible micro-scale filling factor less than unity, unless plasma inhomogeneities move at different speeds.

\subsection{Plume contribution to the polar wind}
\label{sec:results_contr}

The contribution of plumes to the polar wind with respect to interplumes can be estimated as the ratio of the mass flux due to plumes to the mass flux due to interplumes,

\begin{equation}
  C(r)=\frac{n_{\rm e,pl}(r)w_{pl}(r)ff(r)}{n_{\rm e,ip}(r)w_{ip}(r)}
,\end{equation}

where $ff$ is the filling factor of a single plume at the coronal altitude $r$. We implicitly assume that the parts of the coronal hole that have not been recognized as plumes, must be considered as interplumes.

We used the following linear fits to the interplume outflow speeds for O~{\sc{vi}} and protons, in order to calculate the interplume plasma flux in the polar wind. We expect that the validity of the linear fits will be restricted to the heliocentric distances explored by the present study:

\begin{equation}
  w_{\rm O^{+5}}(r)=-344.02+254.88r
  \label{eq:voviipfit}
,\end{equation}

\begin{equation}
  w_{\rm H~Ly\alpha}(r)=-139.48+124.145r
  \label{eq:vpipfit}
.\end{equation}

\begin{table}
  \caption{Polar wind percentage contribution for each single plume, estimated
    from the DD analysis on O~{\sc{vi}} and H~{\sc{i}} Lyman~$\alpha$
    intensities.}
\centering
\begin{tabular}{l r r}
\hline\hline
r  & \% O VI  & \% H Lya \\
\hline
Plume 0 \\
\hline
 1.45 & $ 5.4^{+4.0}_{-0.6}$ & $ 3.3^{+2.2}_{-1.9}$ \\
 1.52 & $ 2.1^{+1.7}_{-0.9}$ & $ 3.2^{+1.7}_{-1.4}$ \\
 1.60 & $ 5.9^{+2.3}_{-1.7}$ & $ 3.8^{+1.3}_{-1.2}$ \\
 1.70 & $ 5.2^{+2.3}_{-1.5}$ & $ 3.7^{+1.2}_{-1.1}$ \\
 1.88 & $12.1^{+4.4}_{-3.0}$ & $10.0^{+2.6}_{-2.2}$ \\
 2.21 & $ 2.9^{+2.2}_{-1.3}$ & $ 4.5^{+0.9}_{-2.8}$ \\
\hline
Plume 1 \\
\hline
 1.41 & $ 2.8^{+4.8}_{-2.8}$ & $ 1.5^{+2.4}_{-1.4}$ \\
 1.47 & $ 1.9^{+2.0}_{-1.9}$ & $ 1.6^{+1.6}_{-1.6}$ \\
 1.56 & $ 3.7^{+1.7}_{-1.5}$ & $ 3.7^{+1.4}_{-1.4}$ \\
 1.65 & $ 3.3^{+1.5}_{-1.5}$ & $ 3.5^{+1.6}_{-1.5}$ \\
 1.77 & $ 7.8^{+1.5}_{-1.5}$ & $ 7.5^{+1.2}_{-1.2}$ \\
\hline
Plume 2 \\
\hline
 1.53 & $ 5.1^{+2.3}_{-2.3}$ & $ 2.7^{+2.1}_{-2.0}$ \\
 1.63 & $ 7.9^{+2.6}_{-2.4}$ & $ 8.2^{+2.2}_{-2.4}$ \\
 1.74 & $ 3.6^{+2.3}_{-2.2}$ & $ 3.8^{+2.3}_{-2.3}$ \\
 1.86 & $ 8.4^{+6.4}_{-4.6}$ & $ 8.5^{+5.2}_{-4.5}$ \\
\hline
Plume 3 \\
\hline
 1.66 & $ 5.5^{+2.3}_{-2.5}$ & $ 5.1^{+2.0}_{-2.3}$ \\
 1.77 & $ 4.1^{+2.0}_{-2.4}$ & $ 3.5^{+1.8}_{-2.0}$ \\
 1.90 & $ 4.1^{+1.9}_{-1.8}$ & $ 3.9^{+1.9}_{-1.7}$ \\
\hline
\end{tabular}
\label{tab:04}
\end{table}

The results for every single plume are given in Table~\ref{tab:04}, where the percentage contributions are reported for different heliocentric distances, as mean values with errors, for O~{\sc{vi}} and protons.

A possible way of estimating the total plume contribution to the polar wind, $C_{\rm tot}$, is to sum up the mean values of the contribution for all the plumes given in Table~\ref{tab:04}. In the case of O~{\sc{vi}} the result is $C_{\rm tot}=20\substack{+10\\-7}$ percent, and for protons  it is $C_{\rm tot}=18\substack{+8\\-7}$ percent, with the two values fairly compatible within uncertainties.

\subsection{Geometry of plumes and mass flux conservation}
\label{sec:results_geom}

\begin{figure*}
  \centering
  \includegraphics[width=16.0cm]{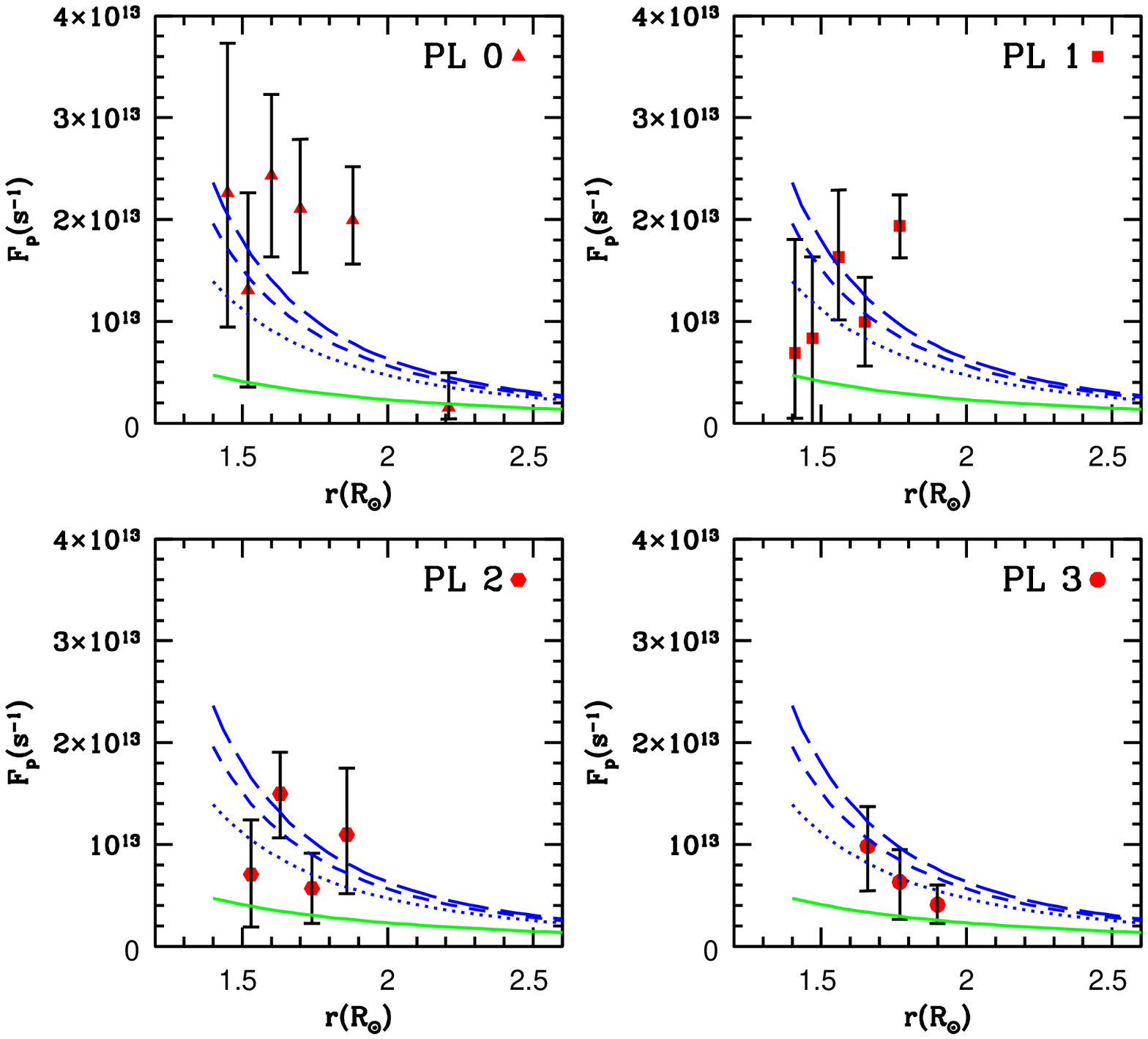}
  \caption{Proton number flux for the single plume model. Solid lines represents the mass flux conservation for a radial expansion; dotted, short dashed, and long dashed profiles describe the super-radial expansion for the magnetic field models of \cite{Banaszkiewicz1998a} with $Q=0.3$, $Q=1.0,$ and $Q=1.5$, respectively.}
  \label{fig:cons}
\end{figure*}

The consistency of the DD analysis with the proton flux conservation can be verified by comparing the proton flux $w_p(r)n_p(r)$, as derived from the DD analysis, with the flux derived from Equation~(\ref{eq:mass_cons}), expected on the basis of the plume geometry, with the boundary condition of the flux measured at 1 AU (see Section \ref{sec:results_geom}):
\begin{equation}
\nonumber
F_p(r)=\frac{F_p(1AU)(215R_{sun})^2}{r^2}\frac{f(1AU,\theta)}{f(r,\theta)}
.\end{equation}
The comparison is shown in Figure~\ref{fig:cons}, where the expected profiles for radial expansion, and super-radial expansion described by the magnetic field model by \cite{Banaszkiewicz1998a}, are shown for three different values of the parameter $Q$. Although the uncertainties from the DD analysis are large, and the fit of a model to the plume profiles does not allow us to establish a well defined geometry, the major part of the experimental points are compatible with the set of plume profile geometries we inferred from LASCO, UVCS, and EIT images, and with the condition of mass flux conservation, as in the case of plumes PL2 and PL3. In the case of plumes $P_0$ and $P_1$, the results apparently contradict the mass flux conservation for any value of the expansion factor.

\section{Discussion and conclusions}
\label{sec:discussion}

The purpose of the present paper was to provide a contribution to the discussion on whether plumes are to be considered dynamic structures or not, and in the affirmative case, to help in quantifying their contribution to the solar polar wind. The results we obtained confirm that plumes are substantially denser than the interplumes, and the density contrast turns out to be about 3-5, depending on the altitude in the corona.  The outflow speeds suggest that plumes are not static structures, at least for distances higher than about 1.6\rsun. In some cases, for example, for plumes PL0 and PL1, the speed profiles appear similar to the profiles typical of interplumes. In other cases, such as those of plumes PL2 and PL3, the outflow speeds differ significantly from those of interplumes, being characterized by a lower acceleration with distance. The electron density and outflow speed values in plumes we derived from the DD analysis, and the estimated plume filling factor in the polar coronal hole led us to evaluate the contribution of plumes to the polar wind as being about 20\% of the mass flux, and this result should be considered as a lower limit (see discussion in Section \ref{sec:results_contr})

The results we obtained for different plumes are not uniform, in the sense that sometimes we derive low speed values, as we would expect on the basis of previous results published in the literature (see \citealt{Wilhelm2011a}, \citealt{Poletto2015a}). Sometimes the outflow speeds do not differ much from the interplume values, and we would like to comment upon this. Signatures of outflows in plumes have been found in the low corona from observations taken with SoHO/SUMER and Hinode/EIS (see \citealt{Wilhelm2000a}, \citealt{Tian2010a}). More recently, a study based on SDO/AIA (Solar Dynamics Observatory/Atmospheric Imaging Assembly) data published by \cite{Pucci2014a} analyzed the propagation of
disturbances in a polar coronal hole in the low corona, during an entire plume lifetime of about 40 hrs. They concluded that they represent high-speed outflowing events, with distributions peaking at $167~{\rm km~s^{-1}}$ and $100~{\rm km~s^{-1}}$ for plumes and interplumes, respectively.

On the basis of the model of polar plume formation proposed by \cite{Wang1995a}, which is characterized by the reconnection of closed bipolar regions with open unipolar magnetic flux tubes in polar coronal holes, the results obtained in the low corona suggest a scenario in which the outflow in a coronal hole is characterized by many episodes, which are more frequent in connection with plumes and evolve during the lifetime of a plume. Hence, in the plumes in the intermediate corona we could expect to observe an episodic outflow speed regime, which can not be resolved in time by UVCS observations, and the results of the DD analysis should be regarded as representative of an average (see Section \ref{sec:results_dd}). We may also speculate that the high and low speed regimes we observed in different plumes correspond to different stages in the evolution of a plume. Moreover, a variable and evolving outflow speed in plumes could also explain why, in some cases, we found that the mass flux conservation constraint along plumes is apparently not respected by the flux estimates we measured at different altitudes by applying the DD technique (see discussion in Section \ref{sec:results_geom}).

The analysis presented in this paper is based on a single case study, hence we do not know how much it could be representative of the real dynamics in plumes. The analysis of a larger sample of UVCS observations will be the subject of a forthcoming study, where we plan to investigate the effects of a possible speed regime evolution during a plume life time.

A new opportunity will become available with the Metis coronagraph on-board the Solar Orbiter spacecraft (see \citealt{Antonucci2017a}; \citealt{antonucci2019a}). Metis will provide simultaneous observations of the solar corona in visible light and H~{\sc{i}}~Lyman~$\alpha$, with the reasonably high cadence of 20 min., over time intervals of ten days during each perihelion. By means of the polarimetric analysis of visible light emission and the DD technique applied to the H~{\sc{i}}~Lyman~$\alpha$, it will be possible to investigate in detail the evolution of individual polar plumes and to establish more firmly their contribution to the slow solar wind.

\begin{acknowledgements}
The authors would like to thank Giannina Poletto for having read the manuscript and for the many helpful comments and suggestions, and the anonymous referee for the review of our paper.
\end{acknowledgements}

\bibliographystyle{aa} 

\end{document}